\documentclass[prx,reprint,aps,amsmath,amssymb,amsfonts,superscriptaddress,notitlepage,floatfix]{revtex4-1}

\usepackage[colorlinks=true,citecolor=blue,urlcolor=blue,linkcolor=blue,hyperfigures=true]{hyperref}

\usepackage[utf8]{inputenc}
\usepackage[T1]{fontenc} 

\usepackage{graphicx}
\usepackage{dcolumn}
\usepackage{bm}
\usepackage{braket}
\usepackage{color}
\usepackage{multirow}
\usepackage[percent]{overpic}
\usepackage{changes}
\usepackage{bbold}
\usepackage{amsmath}
\usepackage{gensymb} 
\usepackage{changes}
\usepackage{tabularx} 

\begin{document}

\title{Plasmonic Quantum Dots in Twisted Bilayer Graphene}

\author{T. Westerhout}
\author{M. I. Katsnelson}
\author{M. R\"osner}
\email{m.roesner@science.ru.nl}
\affiliation{Institute for Molecules and Materials, Radboud University, Heijendaalseweg  135, 6525 AJ Nijmegen, The Netherlands}

\date{\today}

\begin{abstract}

    We derive a material-realistic real-space many-body Hamiltonian for twisted bilayer graphene from first principles, including both single-particle hopping terms for $p_z$ electrons and long-range Coulomb interactions. By disentangling low- and high-energy subspaces of the electronic dispersion, we are able to utilize state-of-the-art constrained Random Phase Approximation calculations to reliably describe the non-local background screening from the high-energy $s$, $p_x$, and $p_y$ electron states for arbitrary twist angles. The twist-dependent low-energy screening from $p_z$ states is subsequently added to obtain a full screening model. We use this approach to study real-space plasmonic patterns in electron-doped twisted bilayer graphene supercells and find, next to classical dipole-like modes, also twist-angle-dependent plasmonic quantum-dot-like excitations with $s$ and $p$ symmetries. Based on their inter-layer charge modulations and their footprints in the electron energy loss spectrum, we can classify these modes into ``bright'' and ``dark'' states, which show different dependencies on the twist angle. 
    
\end{abstract}

\maketitle

\section{Introduction}

    Layered materials with weak inter-layer van der Waals (vdW) interactions allow for precise control of the inter-layer twist angle. The resulting moir\'e potential has been shown to yield fascinating effects. For example, in the case of twisted bilayer graphene a small ``magic angle'' has been theoretically predicted at which ultra-flat bands form~\cite{suarez_morell_flat_2010,bistritzer_moire_2011}. Together with sizable Coulomb interactions this allows for possibly strong correlation effects. Both, the characteristics of ultra-flat bands~\cite{li_observation_2010} as well as correlation effects have been experimentally verified in the form of insulating and superconducting gaps as well as in the form of ferromagnetic behavior controlled by the doping level~\cite{cao_correlated_2018,cao_unconventional_2018,yankowitz_tuning_2019,sharpe_emergent_2019}. For twisted semiconducting layered materials, such as transition metal dichalcogenides (TMDCs), the effects of the moir\'e potential on the excitonic properties have been theoretically \cite{wu_topological_2017,danovich_localized_2018,wu_hubbard_2018,brem_tunable_2020,choi_twist_2021} and experimentally \cite{tran_evidence_2019,seyler_signatures_2019,zhang_twist-angle_2020} studied. In these cases, a finite twist angle can yield superlattices with periodicities of the order of the excitonic radii, which can again yield flat electronic dispersions~\cite{li_imaging_2021,angeli__2021} and can effectively trap exciton complexes~\cite{seyler_signatures_2019,brem_tunable_2020}. 
    
    For both, correlation effects in twisted bilayer graphene as well as for the formation of moir\'e excitons in twisted TMDCs, the Coulomb interaction plays a major role. While the single-particle properties of these twisted materials have been studied in great detail including the ab initio derivation of the moir\'e potentials, the Coulomb interaction has so far been treated with less care. 
    For twisted bilayer graphene various models have been utilized ranging from purely local Coulomb interactions~\cite{cao_correlated_2018}, to non-local interactions taking the effective thickness and/or the dielectric environment into account\cite{goodwin_twist-angle_2019,cea_band_2020,bernevig_twisted_2021,liu_tuning_2021}, and to models treating the low-energy $p_z$ screening on the level of the (constrained) Random Phase Approximation~\cite{pizarro_internal_2019,goodwin_attractive_2019}.
    To describe the Coulomb interaction in twisted bilayer TMDCs various models have been suggested and used including models resolving the intra- and inter-layer Coulomb interactions based on ab initio estimates for the relevant dielectric functions or constants~\cite{danovich_localized_2018,brem_tunable_2020,choi_twist_2021}.
    
    Here, we go beyond these effective Coulomb descriptions by deriving an interacting low-energy model for twisted bilayer graphene including both single-particle and Coulomb interaction matrix elements via state-of-the-art down folding of ab initio calculations. With this we especially aim to consistently describe the screening from low-energy $p_z$ orbitals \emph{and} from the remaining bands, i.e. from $sp^2$ and all other higher-energy states. 
    We use this model to calculate plasmonic excitations in real space for electron-doped finite-size samples at moderate and large twist angles. So far, plasmons in twisted bilayer graphene have been mostly investigated with a focus on small (``magic'') angles~\cite{sunku_photonic_2018,lewandowski_intrinsically_2019,hesp_collective_2019,novelli_optical_2020} with the exception of Ref.~\cite{hu_real-space_2017}. 
    Our focus on larger angles comes with a major methodological advantage, which is the circumvention of the screening properties from extremely flat bands and thus large density of states, as it is in the latter situation still not entirely clear whether the Random Phase Approximation is applicable~\cite{katsnelson_anomalies_1985,irkhin_robustness_2002,stepanov_coexisting_2021}.
    Additionally, the inter-layer hopping model for larger twist angles is less delicate than for small angles. In the latter case one needs to take into account atomic relaxations numerically~\cite{shi_large-area_2020} or at a model level by introducing a family of topological defects~\cite{gornostyrev_origin_2020}, which makes the calculations challenging. For moderate to large twist angles, atomic relaxation effects turn out to be negligible, as was shown in Ref.~\cite{wijk_relaxation_2015} using atomistic simulations with realistic carbon potentials. The use of a nominal, purely geometric moir{\'e} structure is therefore justified for larger twist angles. 
    
    Within the outlined framework, circumventing common modeling issues, we are able to reliably describe real-space plasmonic excitations from first principles at various doping levels for the case of moderate and large twist angles. Next to a variety of conventional dipole and multipole plasmonic patterns we find strongly extended patterns with $s$ and $p$-like symmetries, which we identify as plasmonic quantum dot states.
    Due to the layered structure of our material, we find $s$ wave plasmonic quantum dot states with in- and out-off-phase charge accumulation with respect to the layer. Their corresponding footprints in the electron energy loss spectra are very different, and we identify them as ``bright'' and ``dark'' plasmonic excitations. We find that the bright $s$-like excitation energy is nearly independent of the twist angle, while the dark one shows a significant reduction of its excitation energy upon a twist by $10^\circ$.

\section{Modeling Approach}

    \subsection{Hamiltonian and Plasmonic Properties}
    
        \begin{figure}
            \includegraphics[width=0.99\columnwidth]{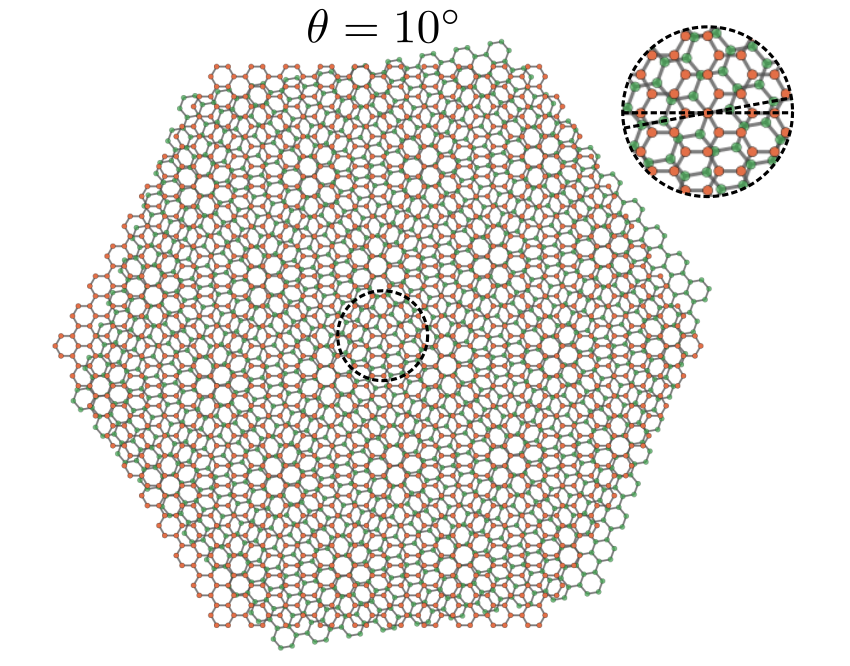}
            \caption{Sketch of a twisted bilayer supercell with 3252 atoms, armchair edges, and at $\theta = 10\degree$. The rotation axis is centered on the A (B) sublattice of the upper (lower) layer.  \label{fig-sc} }
        \end{figure}
    
        We aim to study twisted bilayer graphene supercells as depicted in Fig.~\ref{fig-sc}. These are constructed such that for zero twist angle ($\theta=0\degree$) we get an AB (Bernal) stacked bilayer graphene supercell. The rotation axis is centered at the upper A sublattice (lower B sublattice), as indicated in the inset of Fig.~\ref{fig-sc}. The outer boundaries are chosen to be of armchair type.
        We describe these supercells with a low-energy Hamiltonian for the $p_z$ states
        \begin{align}
            H = \sum_{i,j} t_{ij} c_i^\dagger c_j
              + \frac{1}{2} 
                \sum_{i,j} U_{ij} n_i n_j \,,\label{eq:hamiltonian}
        \end{align}
        with $i$ and $j$ being atomic lattice positions. $c_i$ ($c_i^\dagger$) and $n_i = c_i^\dagger c_i$ are $p_z$-orbital annihilation (creation) and corresponding orbital occupation number operators. $t_{ij}$ and $U_{ij}$ are hopping and density-density Coulomb interaction matrix elements, respectively. At this stage we do not explicitly differentiate between the upper and lower layer in a sense that $i$ and $j$ run over both layers. 
        
        To study plasmonic properties we utilize a real-space version of the Random Phase Approximation (RPA)~\cite{vonsovsky_quantum_1989,giuliani_quantum_2005,wang_plasmonic_2015,westerhout_plasmon_2018,jiang_plasmonic_2021} to calculate the Electron Energy Loss Spectra (EELS) defined by
        \begin{align}
            \operatorname{EELS}(\omega) = 
            -\operatorname{Im}\left[ \frac{1}{\varepsilon_1(\omega)} \right]
        \end{align}
        with $\varepsilon_1(\omega)$ being the ``leading'' eigenvalue (with the largest contribution to EELS) of the full dielectric function
        \begin{align}
            \varepsilon(\omega) = \sum_n \varepsilon_n(\omega) \ket{\phi_n(\omega)}\bra{\phi_n(\omega)} \,.
        \end{align}
        Here $\phi_n(r, \omega) = \braket{r | \phi_n(\omega)}$ is the corresponding eigenvector in real space which renders the plasmonic excitation pattern. Within a real-space tight-binding approximation the RPA dielectric matrix is given by
        \begin{align}
            \varepsilon_{ij}(\omega) =
                \delta_{ij} -
                \sum_{k} U_{ik} \Pi_{kj}(\omega) \label{eqn-Eps_pz}
        \end{align}
        with the Coulomb interaction $U_{ik}$ entering the Hamiltonian \eqref{eq:hamiltonian} and the polarizability function $\Pi_{ij}$ given by
        \begin{align}
            \Pi_{ij}(\omega) = 
                2 \cdot \sum_{ab} \psi_{ia}^* \psi_{ib} \psi_{ja}^* \psi_{jb}
                \frac{f_a - f_b}{E_a - E_b + \omega + i\eta} \,. \label{eqn-Pi_pz}
        \end{align}
        Here $E_a$, $\psi_{ia} \equiv \psi_{a}(i)$ and $f_a$ are eigenvalues, eigenvectors and corresponding Fermi functions obtained upon diagonalization of the single-particle tight-binding Hamiltonian. $i$, $j$, and $k$ label site indices while $a$ and $b$ label eigenstates of the Hamiltonian. $\eta$ is a small positive constant of the order of $1\,$meV. More details on the initial implementation can be found in Ref.~\cite{westerhout_plasmon_2018}. In section~\ref{sec:app_rpa} we furthermore describe how these real-space RPA calculations can be significantly accelerated by exploiting the sparsity of the involved matrices and making use of modern Graphical Processing Units (GPUs). In the following we derive all the necessary model parameters from ab initio.
    
    \subsection{Ab Initio Down Folding}
    
        \begin{figure}
            \includegraphics[width=0.99\columnwidth]{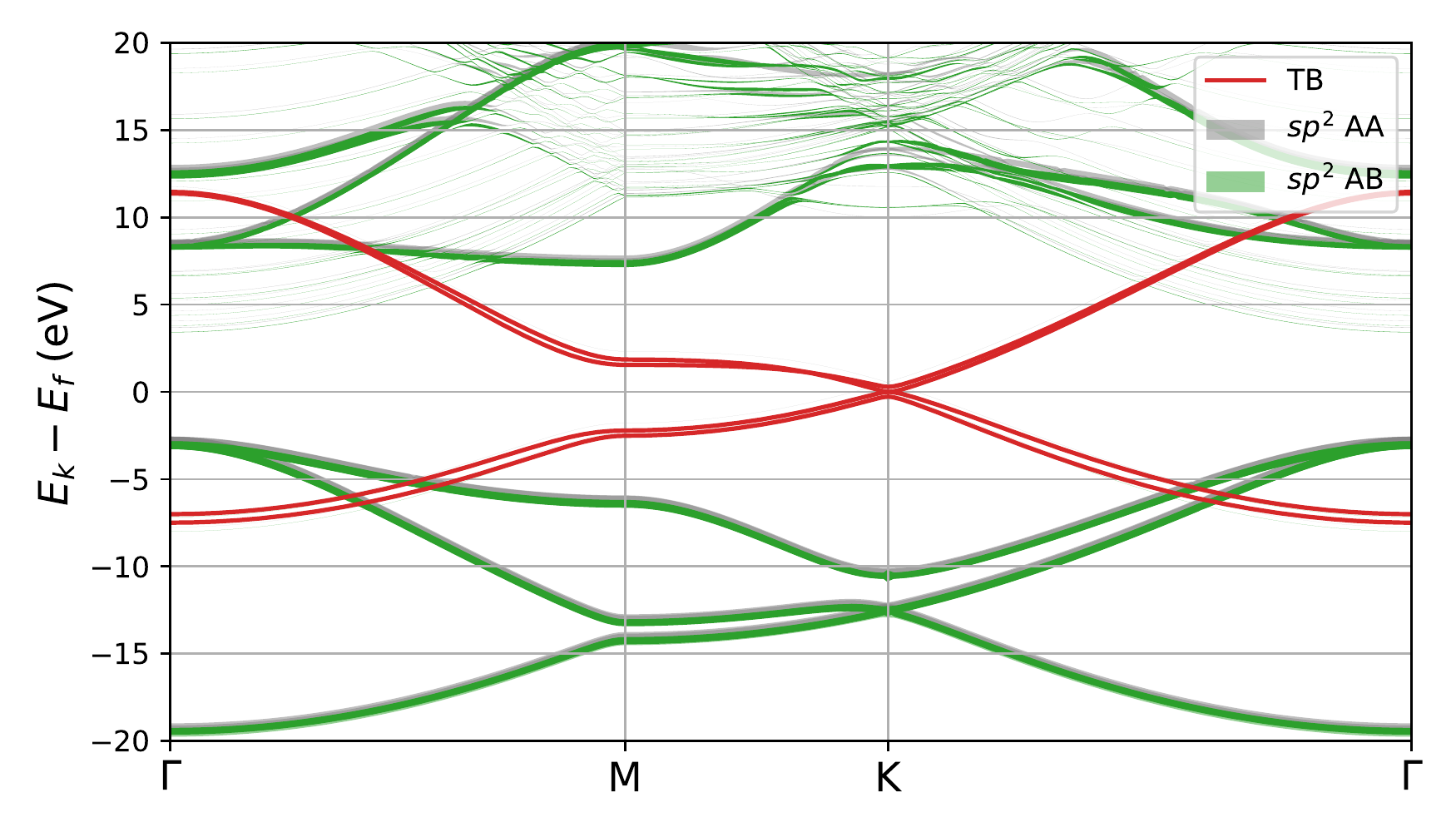}
            \includegraphics[width=0.99\columnwidth]{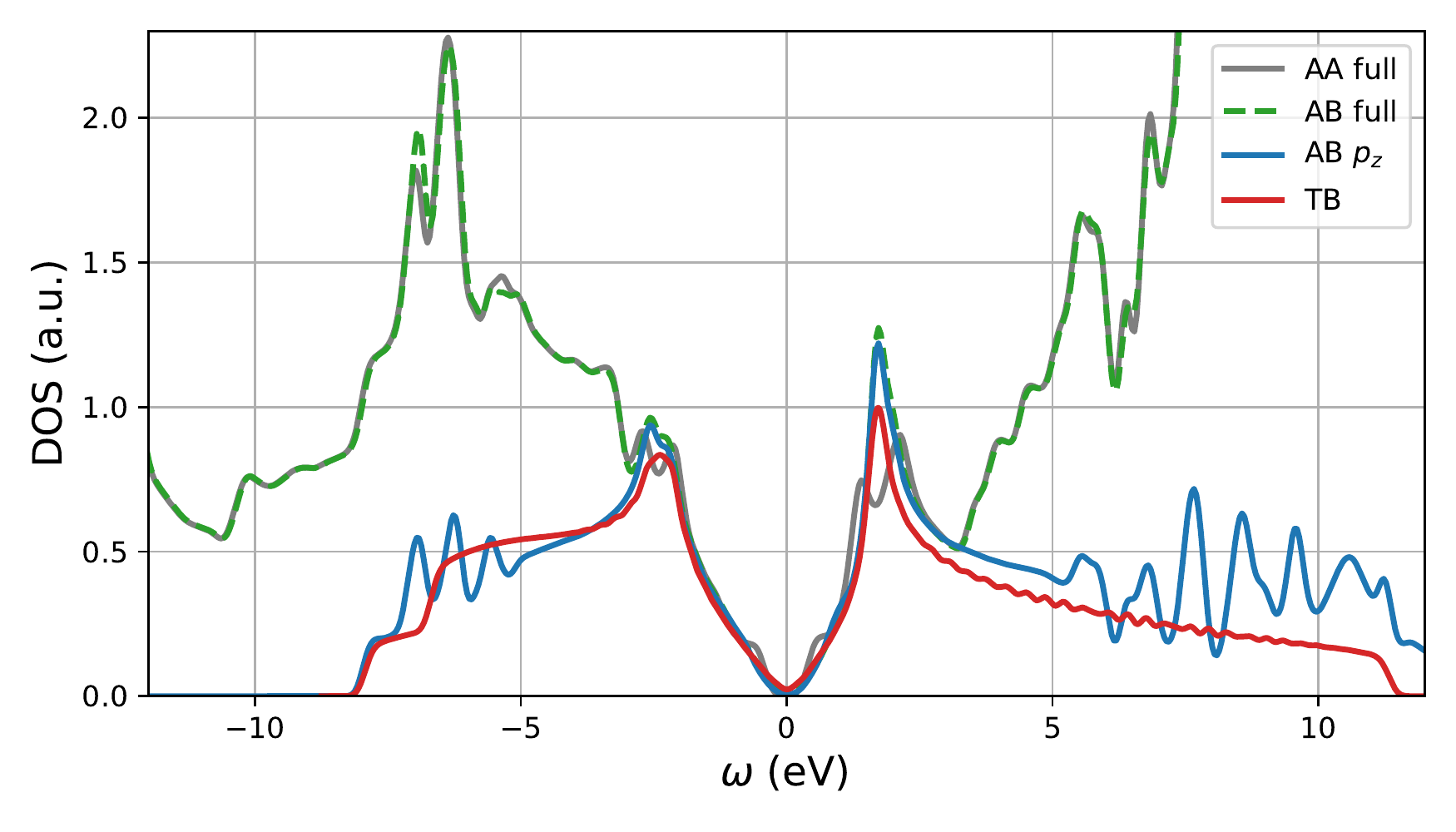}
            \caption{(a) $sp^2$ projected AA (grey) and AB (green) stacked bilayer graphene band structures from DFT together with the Wannier model band structure for the $p_z$ states (red). (b) Full ab initio density of states for AA (grey) and AB (green) stacked bilayer graphene together with the ab initio $p_z$ projection (blue) and the corresponding density of states from our Wannier model, \label{fig-ab} }
        \end{figure}
    
        In order to derive all model parameters for arbitrary twist angles via down folding of first principles calculations we have to make one well justified approximation: only the $p_z$ states will experience the moir\'e potential and thus the twist angle $\theta$. To stress the validity of this assumption we show in Fig.~\ref{fig-ab}~(a) the $sp^2$ projected Density Functional Theory (DFT) band structures of AA (grey) and AB (green) stacked bilayer graphene. Their difference is nearly invisible, as also underlined in Fig.~\ref{fig-ab}~(b) with the comparison of the total density of states (DOS) for these systems. The relative alignment of the two layers thus does not significantly affect the $sp^2$ and higher lying states and can be treated as twist angle-independent.
            
        \begin{table}
            \centering
            \caption{%
                Intra- and interlayer hopping matrix elements for the $p_z$ Wannier orbitals in AB stacked bilayer graphene. Due to the four sublattices we get two different values for the next-nearest-neighbour interlayer hopping.
                \label{tab:hoppings}
            }
            \vspace{0.2cm}
            \begin{tabular}{|c|c|c|c|}
            \hline
            \multicolumn{2}{|c|}{\textbf{Intralayer}} & \multicolumn{2}{c|}{\textbf{Interlayer}} \\
            \hline
            $r$, \AA & $t_{ij}$, eV & $r$, \AA & $t_{ij}$, eV \\ [0.5ex] 
            \hline\hline
            \hspace{0.4cm}0\hspace{0.4cm}
                & \hspace{0.4cm}$-0.991$\hspace{0.4cm}
                & \hspace{0.4cm}3.35\hspace{0.4cm}
                & \hspace{0.4cm}$+0.290$\hspace{0.4cm} \\ 
            1.42 & $-2.857$ & 3.64 & $+0.118$ \\
            2.47 & $+0.244$ & 3.64 & $+0.067$ \\
            2.85 & $-0.258$ & & \\
            3.77 & $+0.024$ & & \\
            4.28 & $+0.052$ & & \\
            4.94 & $-0.021$ & & \\
            5.14 & $-0.014$ & & \\
            5.70 & $-0.022$ & & \\
            \hline
            \end{tabular}
        \end{table}
        
        Based on this assumption we can derive the single-particle hopping matrix elements $t_{ij}$ for the $p_z$ states via a Wannier construction based on DFT calculations for the AB stacked bilayer graphene (see section~\ref{sec:app_ab} for details). In detail, we calculate the hopping matrix elements via
        \begin{align}
            t_{ij}(\theta=0\degree) = \braket{ w_i | H_\text{DFT}^{AB} | w_j }
         \end{align}
         for the untwisted ($\theta=0\degree$) geometry and using $p_z$-like ab initio Wannier functions $w_i(r)$. In Table~\ref{tab:hoppings} we list the resulting intra- and interlayer hopping matrix elements for an interlayer distance of $d=3.35\,$\AA. To account for finite twist angles on the single-particle level, we utilize a Slater-Koster based interlayer hopping model~\cite{guinea_continuum_2019} $t_\perp(r) = \gamma_0 \operatorname{exp}[-\alpha (r-r_0)]$, which we fit to the interlayer hopping matrix elements from Table~\ref{tab:hoppings} and obtain $\gamma_0 = 0.29\,$eV and $\alpha = 5.63\,$\AA$^{-1}$. As mentioned above, for small twist angles one would additionally need to account for modulations in the interlayer distance~\cite{wijk_relaxation_2015,shi_large-area_2020,gornostyrev_origin_2020}, but here we are interested in moderate to large twist angles such that the assumption of a purely nominal twisting is adequate.
         
         The accuracy of our low-energy $p_z$ tight-binding model becomes clear from the comparison to the $p_z$-projected DOS presented in Fig.~\ref{fig-ab}~(b) (also the Wannier band structures interpolates the $p_z$ Kohn-Sham states perfectly around the $K$ and $M$ points, not shown). We find that large twist angles ($\theta > 5\degree$) have only a minor impact on the $p_z$ DOS.
         
\begin{figure}
            \includegraphics[width=0.99\columnwidth]{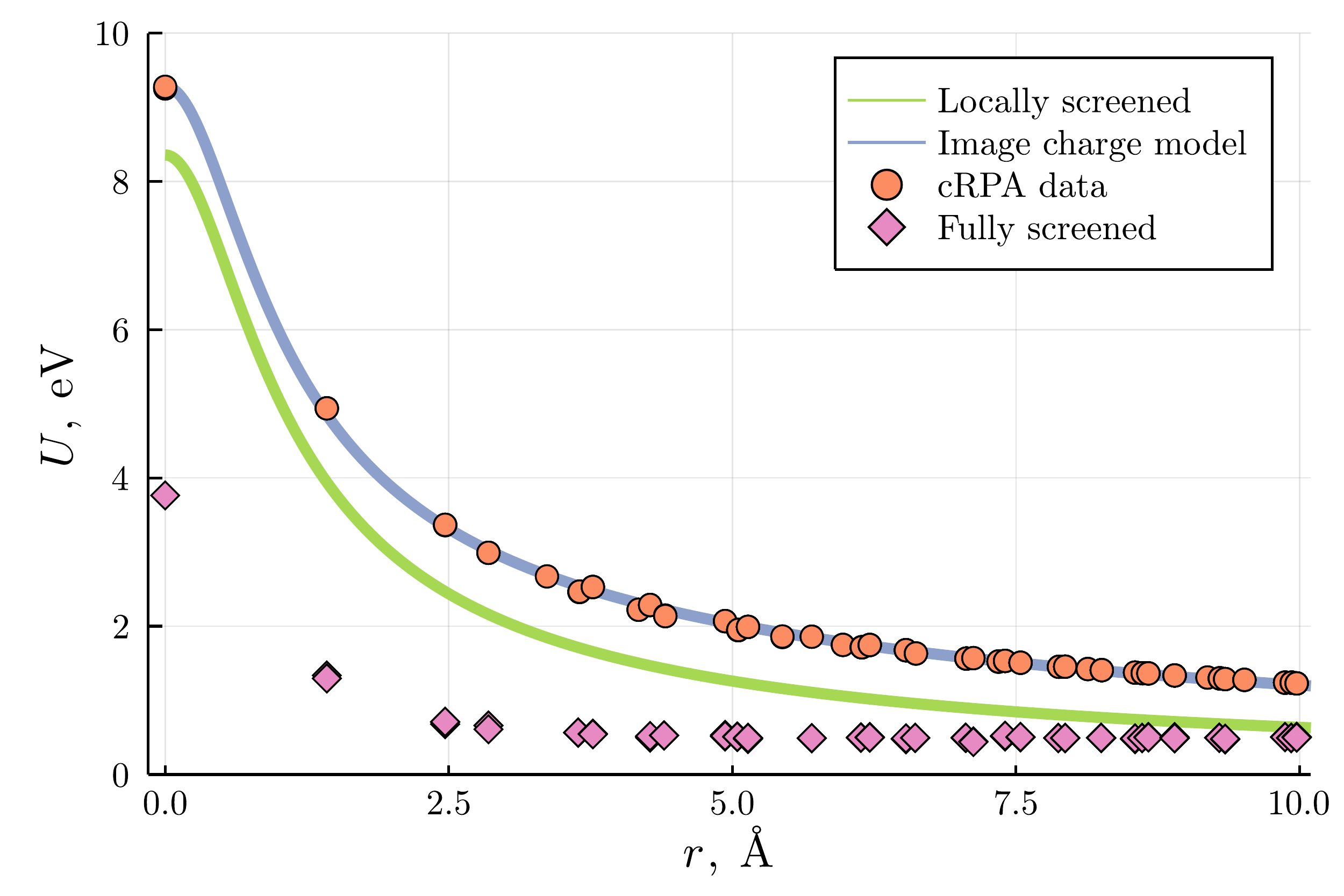}\\
            \hfill\includegraphics[width=0.9\columnwidth]{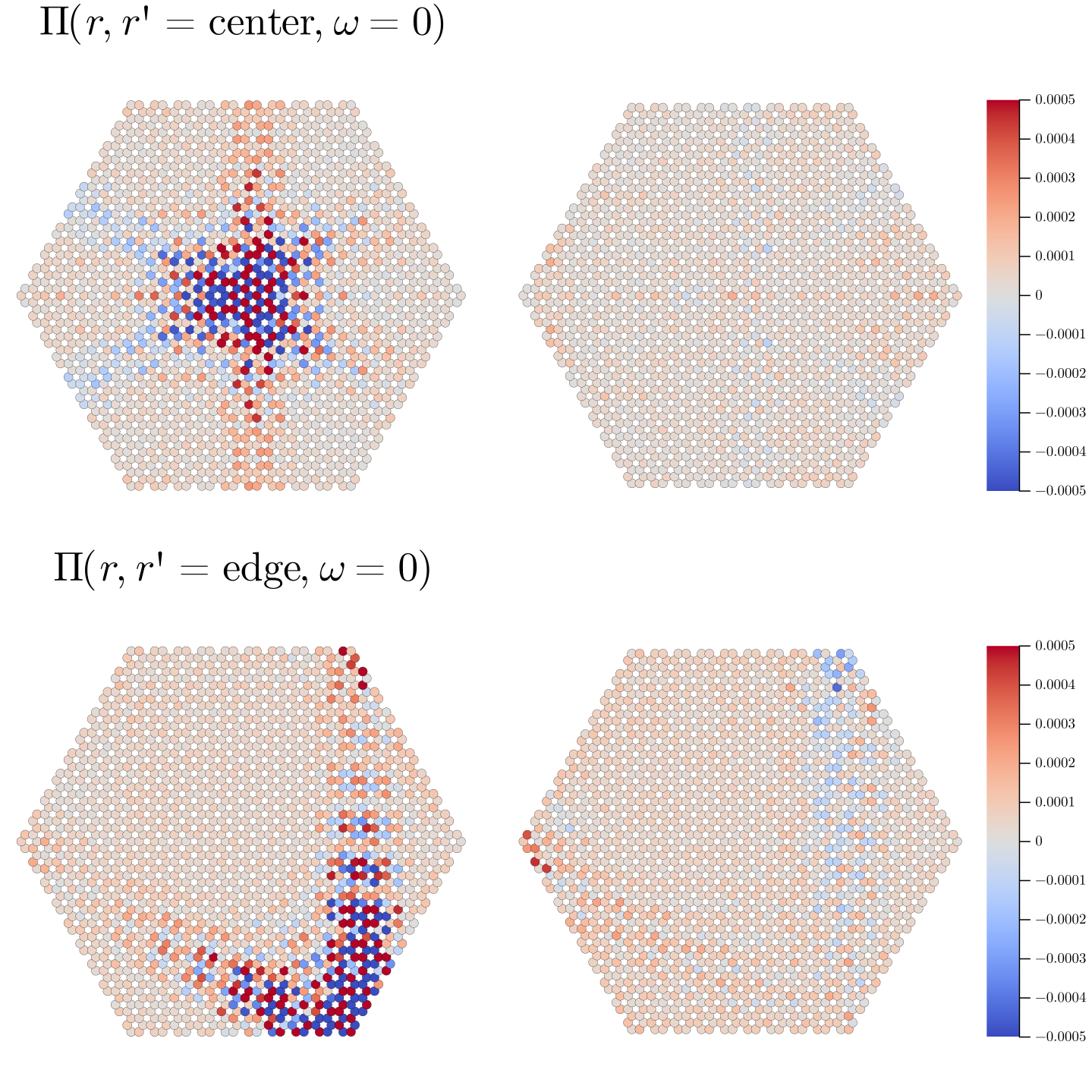}
            \caption{%
                (Top) AB-stacked bilayer graphene Coulomb matrix elements. Orange circles depict cRPA results. The blue line is obtained by fitting Eq.~\eqref{eq:image-charge-model} to the cRPA data. In green we show the locally screened Coulomb interaction (i.e. Eq.~\eqref{eq:image-charge-model} in the limit of $h\to \infty$). Pink diamonds show fully screened Coulomb interaction $W(r,r'=\mathrm{center},\omega=0)$. (Bottom) Static polarizability $\Pi^{p_z}(r, r', \omega=0)$ for two different choices of $r'$. The colorbar is clamped to a small range to highlight the oscillating tails.
                \label{fig-Coulomb}
            }
         \end{figure}

         The description of the fully screened, retarded and $\theta$-dependent Coulomb interaction $W(\omega, \theta)$ requires more attention. $W(\omega, \theta)$ is defined by 
         \begin{align}
             W(\omega, \theta) = \frac{v}{1 - v \Pi^\text{total}(\omega, \theta)} \,,\label{eqn-W_full}
         \end{align}
         where $v$ is the bare Coulomb interaction. $\Pi^\text{total}(\omega, \theta)$ renders all possible screening processes at a given rotation angle $\theta$ which can be separated into two terms:
         \begin{align}
             \Pi^\text{total}(\omega, \theta) \approx 
                 \Pi^{p_z}(\omega, \theta) + \Pi^\text{rest}(\omega = 0),
         \end{align}
         with $\Pi^{p_z}(\omega, \theta)$ being the partial polarization as resulting from virtual excitations within the low-energy $p_z$ sub-space and as defined in Eq.~(\ref{eqn-Pi_pz}). The rest polarization $\Pi^\text{rest}(\omega = 0)$ describes instantaneous screening processes from virtual excitations from and to non-$p_z$ states (such as $sp^2$ and others) as well as ``cross-polarization'' terms from virtual excitations between the two subspaces. Due to the orthogonality of the $p_z$ and $sp^2$ states the cross-polarization terms can be safely neglected~\cite{van_loon_random_2021}. Importantly, this renders the background (or rest) polarization independent of the twist angle. Using this in Eq.~(\ref{eqn-W_full}) yields
         \begin{align}
             W(\omega, \theta) 
             &= \frac{v}{1 - v \left[\Pi^{p_z}(\omega, \theta) + \Pi^\text{rest}(\omega = 0)\right]}\\
             &= \frac{U}{1 - U \Pi^{p_z}(\omega, \theta)} \notag
         \end{align}
         with
         \begin{align}
             U = \frac{v}{1 - v \Pi^\text{rest}(\omega = 0)}
         \end{align}
         being the $\theta$-independent background screened Coulomb interaction, as needed for the evaluation of Eq.~(\ref{eqn-Eps_pz}). We calculate $U$ within the constrained RPA~\cite{cRPA} based on ab initio calculations for AB-stacked bilayer graphene (see section~\ref{sec:app_ab} for details). This yields discretized $U_{ij}$ with $i$, $j$ being AB bilayer graphene lattice positions. For the evaluation of Eq.~\eqref{eqn-Eps_pz} we, however, need to evaluate $U_{ij}$ also for other positions resulting from the finite rotation angles $\theta$. To this end, we map the discretized $U_{ij}$ to a continuum model $U(r = r_i - r_j)$. For the latter we choose the analytic image-charge model for the potential within a dielectric slab of height $d$ reading~\cite{keldysh_coulomb_1979,jena_enhancement_2007,emelyanenko_effect_2008,jiang_plasmonic_2021}:
         \begin{align}
             U(r) = \frac{e^2}{\varepsilon_m z_0(r)}
             + 2 \sum_{n=1}^\infty \frac{e^2 \beta_b^n}{\varepsilon_m z_n(r)}\label{eq:image-charge-model}
         \end{align}
         with $e$ being the elementary charge, $\varepsilon_m$ the dielectric constant of the slab, $z_n(r) = \sqrt{r^2 + \delta^2 + (nh)^2}$, and $\beta_b = (\varepsilon_m - 1) / (\varepsilon_m + 1)$. The additional parameter $\delta$ allows us to also fit the numerical on-site potential $U_{ii} = U(r=0)$. In Fig.~\ref{fig-Coulomb} we show the ab initio cRPA data together with the fit using Eq.~\eqref{eq:image-charge-model}, the locally screened interaction $\frac{e^2}{\varepsilon_m z_0(r)}$ ($h \rightarrow \infty$), and the fully screened interaction $W(\omega=0,\theta=0)$. For the fit we fixed $d = 6.7\,$\AA\ (twice the interlayer distance) and find $\varepsilon_m \approx 2.26$ (in good agreement with similar fits in momentum space~\cite{rosner_wannier_2015}) and $\delta \approx 0.763\,$\AA, which evidently interpolates the ab intio data well. From the comparison to the locally screened interaction we see that the background screening, as described by the second term in Eq.~\eqref{eq:image-charge-model}, acts differently at each $r$ due to its non-local character. The fully screened interaction $W(\omega=0,\theta=0)$ behaves as expected from Thomas-Fermi screening theory in two dimensions~\cite{katsnelson_nonlinear_2006} which predicts a strongly decaying potential with a $r^{-3}$ asymptotic behaviour, but in our calculations we also find a finite offset $c$. This offset $c$ decays with the supercell size and vanishes in the infinite-size limit (not shown). We attribute this behavior to finite-size/boundary effects. In detail, although the polarization function $\Pi^{p^z}(r, r', \omega=0)$ is rather localized, as shown in Fig.~\ref{fig-Coulomb}~(b), it still has some non-vanishing oscillating tails due to the finite Fermi surface. These tails in $r$ are partially missing if $r'$ is fixed to an edge side, which induces the finite offset $c$. Equipped with the continuous representation of the background-screened Coulomb interaction $U(r)$ and the tight-binding model described by the hopping matrix elements from Table~\ref{tab:hoppings} we can evaluate Eq.~\eqref{eqn-Eps_pz} for arbitrary twist angles $\theta$. 
         
\section{Plasmonic Excitations in AB-Stacked Bilayer Graphene Supercells}

    \subsection{Ideal AB stacking}
    
        \begin{figure}
            \includegraphics[width=0.99\columnwidth]{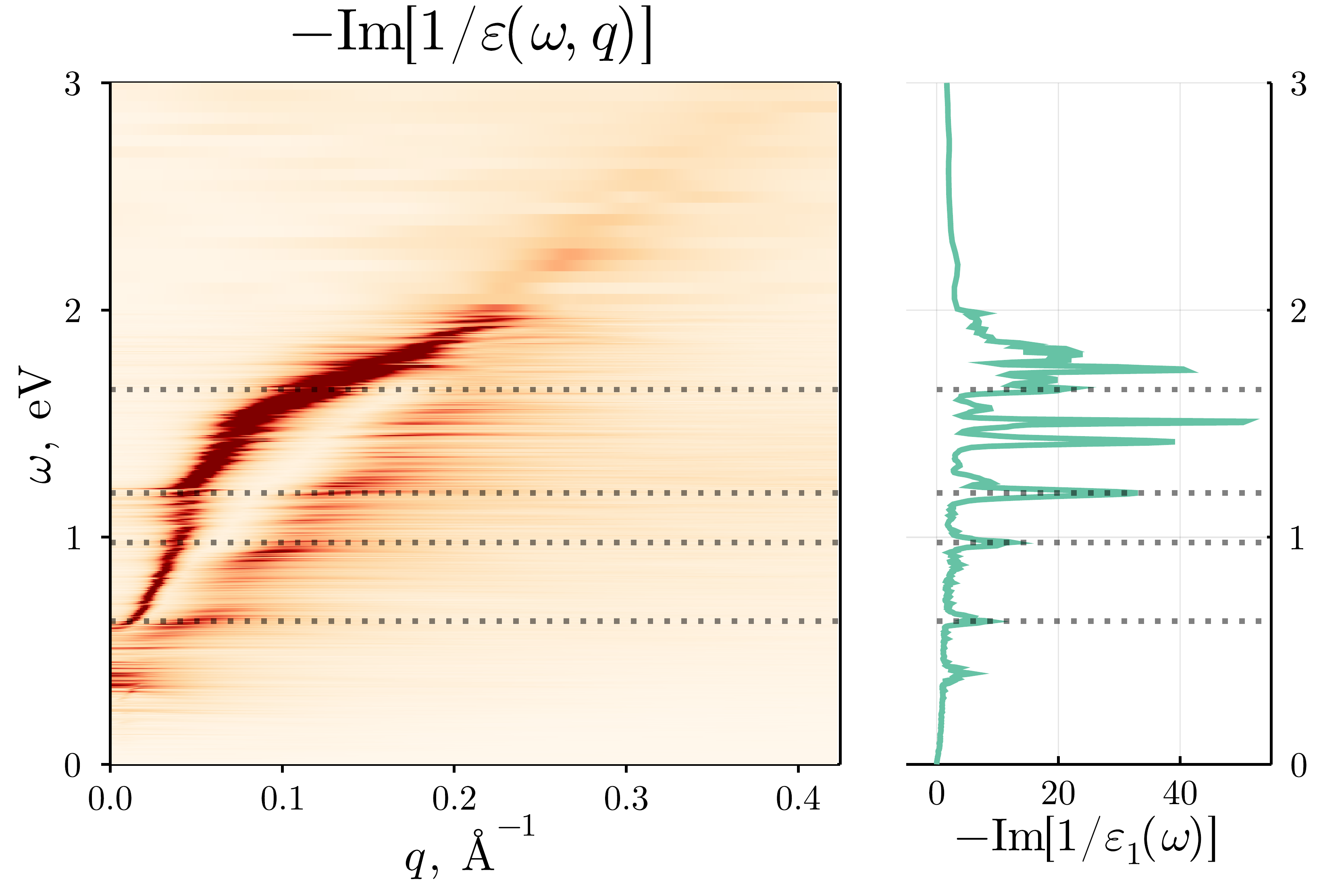}
            \includegraphics[width=0.99\columnwidth]{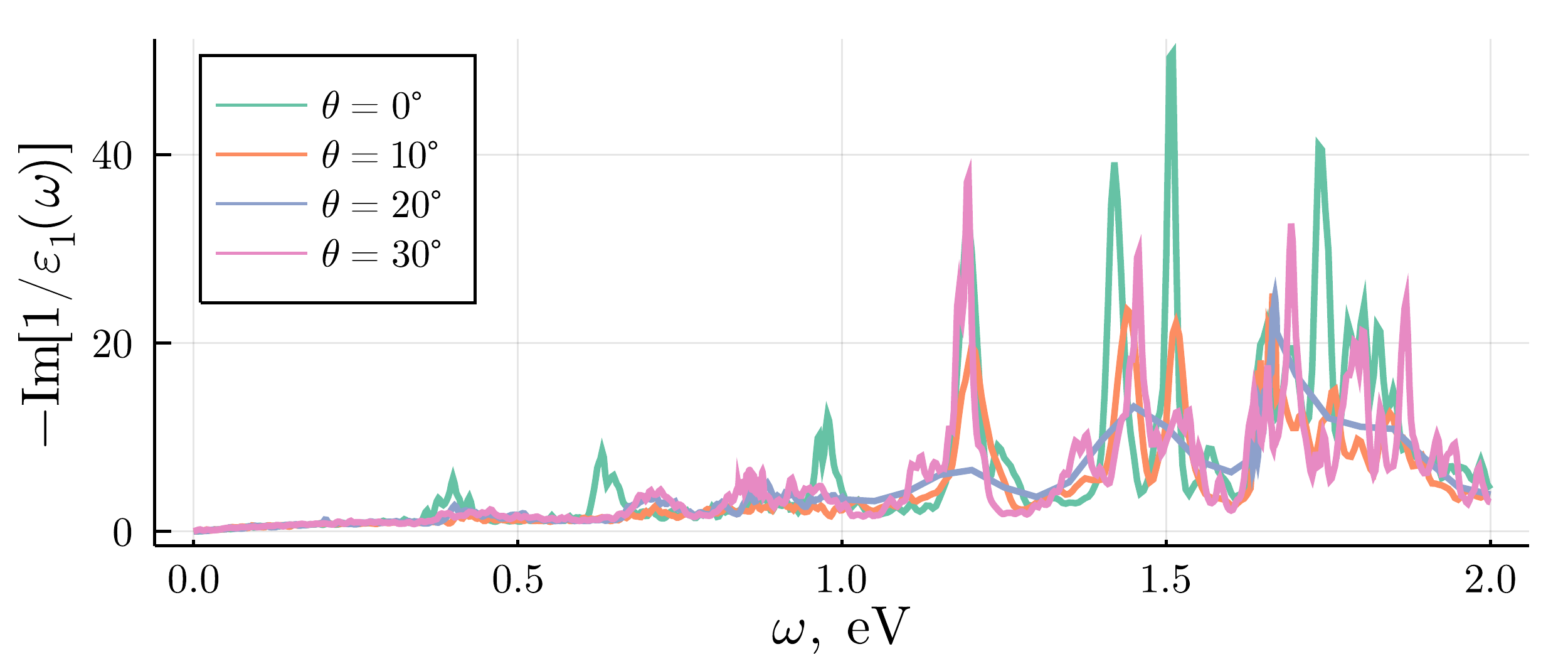}
            \caption{%
                (Top) Plasmonic dispersion relation $\mathrm{EELS}(q, \omega)$ next to full $\mathrm{EELS}(\omega)$ for AB-stacked bilayer graphene (i.e. $\theta = 0\degree$). Dashed lines indicate energies of $0.63$, $0.9775$, $1.195$, and $1.65$ eV corresponding to ``dark'' dipole, ``dark'' 1s, ``bright'' dipole, and ``bright'' 1s modes, respectively. (Bottom) Full $\mathrm{EELS}(\omega)$ for various twist angles $\theta$. \label{fig-EELS}
            }
        \end{figure}
    
        We start our discussion with investigating the EELS of an un-twisted AB stacked bilayer graphene supercell with $3252$ atomic sites corresponding to a side length of $L\approx40\,$\AA\ and at an electron doping of about $n = 5.3\times10^{14}\,$cm$^{-2}$, which is around the maximum achievable with double sided ionic-liquid gating techniques~\cite{zheliuk_josephson_2019}. In Fig.~\ref{fig-EELS} we show the corresponding local EELS($\omega$) next to momentum-resolved EELS($q$, $\omega$) for $\omega < 3\,$eV and for $q<0.4\,$\AA$^{-1}$. In the local EELS we find a series of pronounced excitations below $\omega<2\,$eV. In EELS($q$, $\omega$) there are two plasmonic branches within this energy range. One shows the characteristic ``flattened'' $\sqrt{q}$-like dispersion~\cite{da_jornada_universal_2020,jiang_plasmonic_2021} and the other (lower) one is approximately linear in $q$. Around $\omega=2\,$eV these two modes merge and become strongly Landau-damped yielding a nearly zero EELS($\omega$) for $2\,\text{eV} < \omega < 3\,$eV.

        At this point it is important to note the limits of our material-realistic Coulomb modeling approach. In the ``low-energy'' range ($\omega < 5\,$eV) we are mostly dealing with virtual excitations and thus screening processes solely within the $p_z$ manifold. For larger excitation energies transitions involving the rest space would become important, which we do not correctly render here. Thus, although we find well defined high-energy (gaped) plasmons, we do not discuss them here.
        
        \begin{figure}
            \includegraphics[width=0.99\columnwidth]{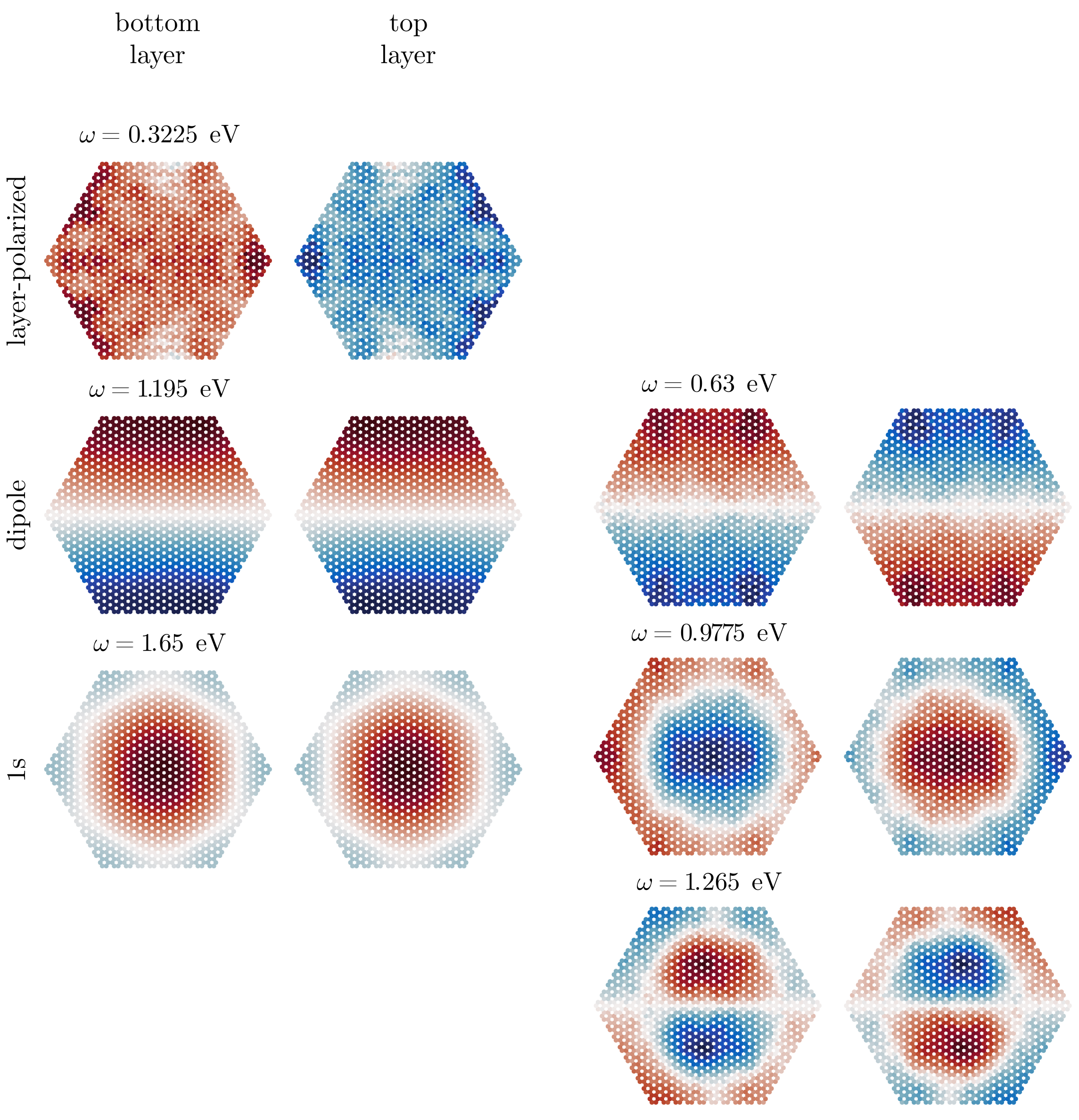}
            \caption{Classical and quantum dot plasmonic modes in real-space for $\theta = 0\degree$ together with their excitation energies. The left (right) two columns depict the bottom and top layer of the ``bright'' (``dark'') modes. \label{fig-modes} }
        \end{figure}
        
        In Fig.~\ref{fig-modes} we categorize the different modes in terms of their real-space charge distributions, as approximately measured by the eigenvector $\phi_1(r, \omega)$ of the dielectric function. For each mode, we show $\phi_1(r, \omega)$ for both layers separately. The first excitation around $\omega \approx 0.32\,$eV is a homogeneous fully layer-polarized mode. Around $\omega \approx 1.2\,$eV we find another charge-polarized mode, but with in-plane dipole character without layer-polarization. This mode is accompanied by a layer-polarized dipole mode at lower frequency $\omega\approx 0.63\,$eV. These extended modes are classically expected for finite size systems and have been found in other 2D systems~\cite{wang_plasmonic_2015}.

        At $\omega \approx 1.65\,$eV we find a quantum-dot like mode with $s$ symmetry without layer polarization and at $\omega \approx 1.0$\,eV its layer-polarized counterpart. In a periodic system we would interpret these modes as the $\omega_+ \propto \sqrt{q}$ and $\omega_- \propto q$ acoustic modes~\cite{das_sarma_plasmons_1998,hwang_plasmon_2009,roldan_dielectric_2013,jin_screening_2015}. Indeed, for $q\approx0.1\,$\AA$^{-1}$ and $\omega \approx 1.65$\,eV and $\omega \approx 1.0$\,eV we find in EELS($q$, $\omega$) two strong resonances in the $\omega_+(q)$ and $\omega_-(q)$ branches, respectively. We can thus interpret these $1s$ plasmonic quantum dot modes as the lowest-energy excitations of this kind in our finite-size supercell. Moreover, we see that the higher excitation energy creates a much stronger EELS signal than the one at lower excitation energy. Based on this observation we refer to them as ``bright'' and ``dark'' $1s$ modes in the following. 
        Like in the case of the classical layer-polarized dipole mode, the layer-polarized ``dark'' $1s$ mode is excited at a lower frequency compared to its ``bright'' counterpart. Due to the interlayer phase shift the total electrostatic/Hartree energy is reduced so that these layer-polarized modes naturally have a lower excitation energy. Next to the $1s$ we also find a ``dark'' (layer-polarized) mode with $p$ symmetry at $\omega \approx 1.27\,$eV. This $p$-like mode, however, cannot be characterized as a conventional extended acoustic mode (i.e. being part of $\omega_{\pm}(q)$), as expected in a periodic system, because of its dipole-like background. In section~\ref{sec:app_doping} we additionally discuss the same modes at lower doping. There, importantly, we loose the ``dark'' modes due to enhanced Landau-damping.
    
    \subsection{Twist angle dependence}
    
         \begin{figure*}
            \includegraphics[width=0.99\textwidth]{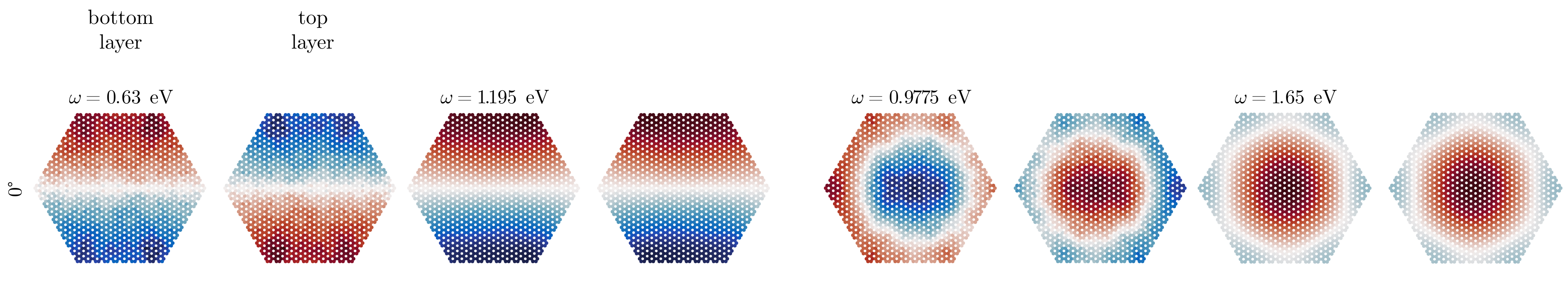}
            \includegraphics[width=0.99\textwidth]{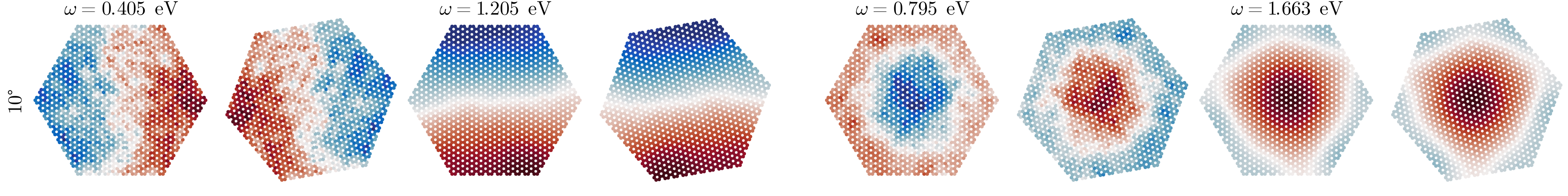}
            \includegraphics[width=0.99\textwidth]{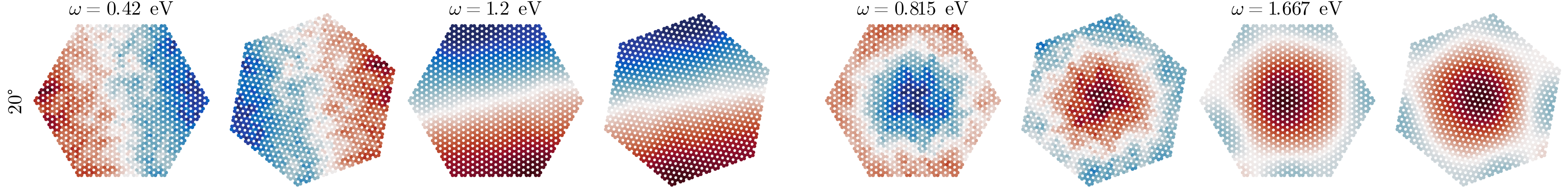}
            \includegraphics[width=0.99\textwidth]{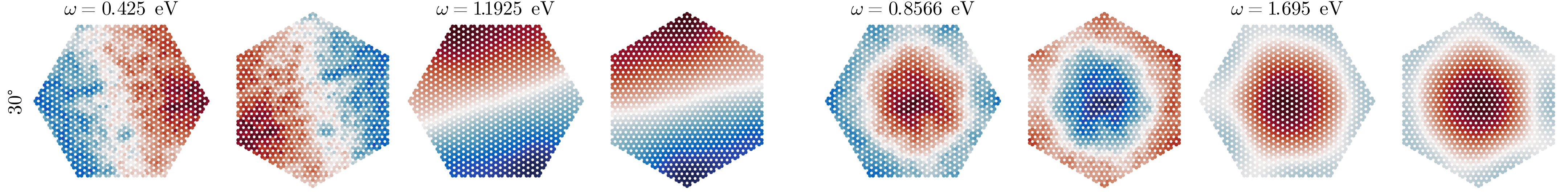}
            \caption{Dipole and $1s$ plasmonic modes in real-space for various $\theta$ together with their excitation energies. The left (right) four columns depict the layer resolved ``dark'' and ``bright'' dipole ($1s$) modes.\label{fig-modesTheta} }
        \end{figure*}
    
        We proceed with the discussion of the impact of finite twist angles on the full EELS as well as on the previously discussed real-space plasmonic patterns and their excitation energies. In the lower panel of Fig.~\ref{fig-EELS} we show EELS($\omega$) for $\theta=0,\, 10,\, 20,\, 30\degree$. From this we see that modes with energies $\omega < 1.0\,$eV are more affected by twisting than the higher-energy excitations. Since it is not clear from this data which mode shifts in which direction, we present in Fig.~\ref{fig-modesTheta} the dark and bright dipole and $1s$ modes for the same $\theta$ together with their corresponding excitation energies. 
        
        For the bright dipole mode we observe that the boundary separating the differently charged areas is rotating when we adjust $\theta$, synchronously in the lower and upper layer, however, only with $\theta/2$. The latter can be readily understood by overlaying the rotated dipole patterns and by remembering that the charge distributions in the two layers are not independent. The missing overlap at the corners of the hexagons additionally yields enhanced charge accumulations in the two opposite corners (per layers). Overall this lowers the mirror symmetry with respect to the charge-separation line to a ``line-inversion'' symmetry. The dark dipole mode behaves similarly. The charge separation line again rotates with $\theta/2$, but we observe a ``smearing'' of it, such that the charge separation is not as sharp as in the bright dipole mode.
    
        In the right two columns of Fig.~\ref{fig-modesTheta} we depict patterns of the $1s$ quantum dot mode. Except for a slight deformation of the initial hexagonal shape we do not see any major changes to the bright excitation. Its dark counterpart also does not show any significant changes in its excitation pattern, except for a smearing of the clear charge separation.
        
         \begin{figure}
			\includegraphics[width=0.99\columnwidth]{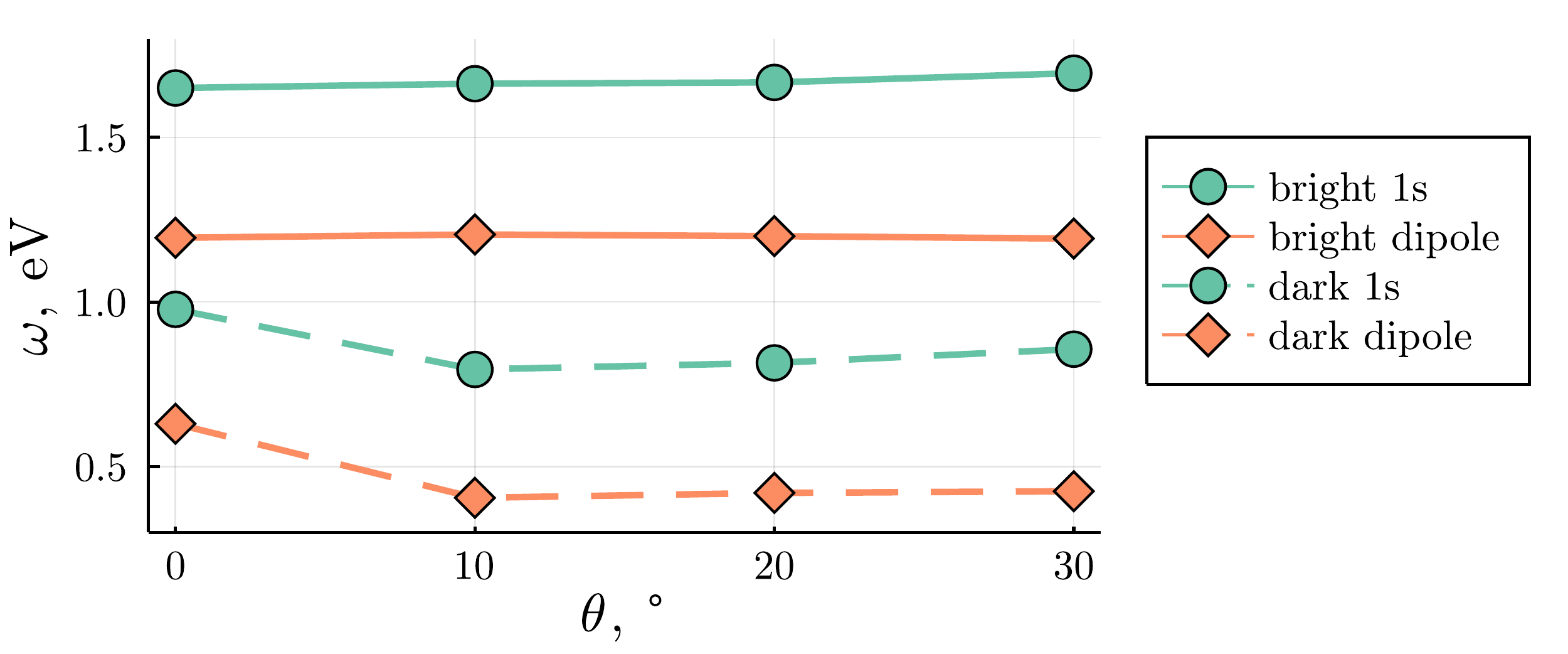}
			\caption{Excitation energies of all modes shown in Fig.~\ref{fig-modesTheta} as a function of $\theta$. \label{fig-ex_vs_deg} }
		\end{figure}
        
        Finally, in Fig.~\ref{fig-ex_vs_deg} we plot the excitation energies for all modes as a function of $\theta$. We see that within the given accuracy the excitation energies of the bright modes do not dependent on the rotation angle. The dark modes, however, do. In detail, we see for both cases, the dark dipole as well as the dark $1s$ mode, a significantly decreased energy upon rotation by $10\degree$. Afterwards, these modes just mildly dependent on further rotation towards $\theta=30\degree$. The initial symmetry breaking between $0\degree$ and $10\degree$ thus leaves the strongest footprint in the plasmonic energies, while the larger rotation angles do not change it too drastically anymore. Due to the lower excitation energy of the dark dipole modes they are energetically closer to the particle-hole continuum and thus more affected by Landau-damping effects. This simultaneously renders these modes also more dependent to the single-particle properties. Since the single-particle properties experience changes upon twisting (due to our $t_\perp$ model), while the Coulomb interaction model is fully rotationally invariant, we understand that the $\theta$-dependent changes to the excitation energy of the dark modes are mostly induced by changes in the single-particle properties.

    \section{Conclusions \& Outlook}
    
        We have presented an ab initio derived twisted bilayer graphene many-body model including a consistent description of the kinetic (hopping) and Coulomb matrix elements. Upon separating the different screening channels into the low-energy $p_z$ and residual background screening, we were able to map the rotation dependence of the total polarizability to the low-energy screening channels only. This allowed us to calculate the background-screened Coulomb interaction from first principles using constrained RPA and to fit the resulting partially-screened interaction with a lightweight continuum model. All rotation dependencies are subsequently handled within the low-energy $p_z$ space only. 
        
        Based on this model we studied low-energy plasmonic excitations in real space in electron-doped twisted bilayer graphene supercells. We observed a variety of different excitation patterns including classical dipole as well as plasmonic quantum dot states. The two layers yield two versions of these excitations: ``bright'' and ``dark'' ones with in- and out-of-phase interlayer charge oscillations. While the bright excitations show no significant twisting dependence, the dark ones show a reduction of their excitation energies upon a twist by $10\degree$. Larger rotation angles change the excitation energies just a bit.
        
        The observed quantum dot plasmonic patterns could be classified in terms of their symmetries into $1s$ and $1p$. These states show promise for both analyzing twisted bilayer systems and for possible practical applications. For example, the twisting dependence of the dark $1s$ mode could be utilized within scanning near-field optical microscopy measurements~\cite{fei_gate-tuning_2012,chen_optical_2012} to measure small (local) twist angle variations. On the other side, the plasmonic quantum dot states might also allow for twist-dependent tailoring of light-matter interactions. The orientation of the $p$-like states might, for example, be utilized to create novel direction-dependent light sensors.
        
        We thus expect that this initial study forms the ground for further material-specific quantitative real-space plasmonics studies of twisted bilayer graphene and similar systems.
     
    \section*{Acknowledgments}
    
        We thank Merzuk Kaltak for sharing his cRPA implementation with us. The work of M.I.K. and T.W. was supported by European Research Council via Synergy Grant 854843-FASTCORR. Numerical simulations in this work were carried out on the Dutch national e-infrastructure with the support of SURF Cooperative.
        
    \section*{Appendix}
    
        \subsection{Doping dependence \label{sec:app_doping}}
        
         \begin{figure}
			\includegraphics[width=0.99\columnwidth]{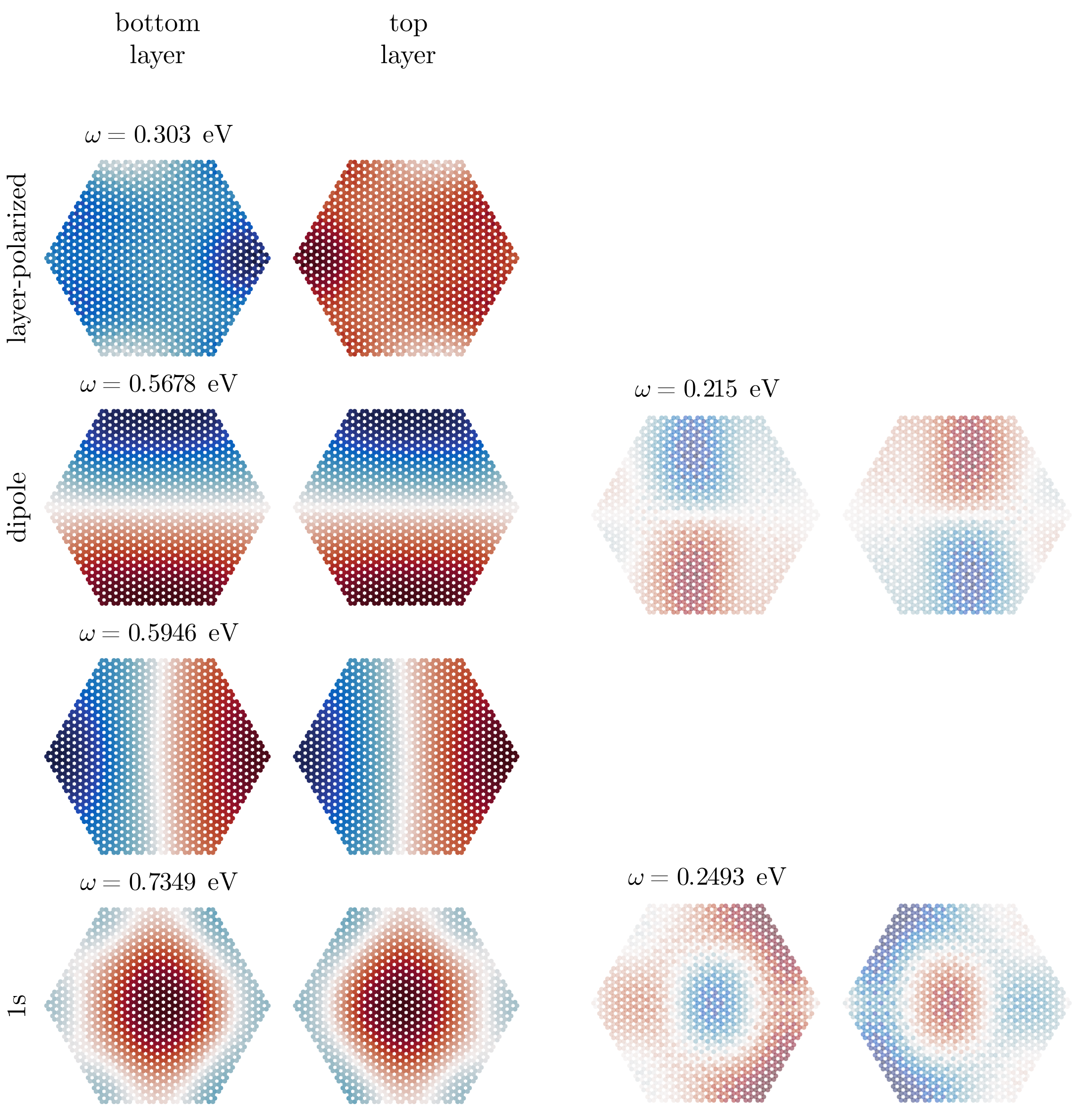}
			\caption{Classical and quantum dot plasmonic modes in real-space for $\theta = 0\degree$ together with their excitation energies are reduced doping. The left (right) two columns depict the bottom and top layer of the ``bright'' (``dark'') modes. \label{fig-doping} }
		\end{figure}
		
		In Fig.~\ref{fig-doping} we present a few plasmonic excitation patterns for a smaller electron doping of $n = 6.3\times10^{13}\,$cm$^{-2}$ and the same supercell as before. In this case we can again clearly identify a variety of bright modes. The dark modes are, however, not well defined anymore. Although we find some plasmonic eigenvectors which resemble the corresponding dark modes at lower excitation energies, these are not well defined plasmonic excitations since the real part of the dielectric function does not fulfill the necessary requirement $\operatorname{Re}\left[ \varepsilon(\omega) \right] = 0$. The excitation energies of the bright dipole and $1s$ modes are strongly reduced compared to the corresponding mode at high electron doping. For the $1s$ mode these observations are fully in line with the expected behaviour of the previously mentioned $\omega_{\pm}(q)$ modes~\cite{das_sarma_plasmons_1998,hwang_plasmon_2009,roldan_dielectric_2013}. The vanishing / fading of the dark modes is thus a result of their close vicinity to the electron-hole continuum.
    
        \subsection{Ab initio details \label{sec:app_ab}}
        
        The band structure and density of states were calculated within density functional theory utilizing the projected augmented wave (PAW) formalism~\cite{paw1, paw2} as implemented in the \emph{Vienna ab initio simulation package} ({\sc vasp})~\cite{Kresse1, Kresse2}. The exchange-correlation effects were considered using the generalized gradient approximation (GGA)~\cite{gga}. A $517\,$eV energy cut-off for the plane-waves and a convergence threshold of $10^{-7}\,$eV were used in the calculations. The Brillouin zone was sampled by a ($18 \times 18$) ${\bf  k}$-point mesh. The in-plane lattice constant is set to $2.468\,$\AA\ and the out-of-plane distance between the two-layer is set to $3.35\,$\AA. A $25\,$~\AA-thick super-cell was used in the direction perpendicular to the 2D plane in order to reduce spurious interactions between supercell images.
        The Wannier functions and the tight-binding Hamiltonian were calculated within the scheme of maximal localization~\cite{mlwf1,mlwf2} using the {\sc wannier90} package~\cite{wannier90}. 
        
        The Coulomb interaction was evaluated using the maximally localized Wannier functions within the constrained Random Phase Approximation (cRPA)~\cite{CoulombU, KaltakcRPA} as ${U_{ij}=\langle w_i w_j |U|w_j w_i\rangle}$, where $U$ is the partially screened Coulomb interaction defined by ${U=v+v \Pi^\text{rest} U}$ with $v$ being the bare Coulomb interaction, $\Pi^\text{rest}$ the cRPA polarization, and $w_i$ is the Wannier function at the lattice site $i$. The polarization operator $\Pi^\text{rest}$ describes screening from all electronic states except those given by the tight-binding Hamiltonian obtained in the Wannier basis. For these calculations, we used a recent cRPA implementation by Kaltak within {\sc vasp}~\cite{KaltakcRPA}. To converge the cRPA polarization with respect to the number of empty states we used in total $208$ bands.
        
        \subsection{RPA details \label{sec:app_rpa}}
        
        The screening from the low-energy $p_z$ orbitals is calculated using the real-space Random Phase Approximation code from Ref.~\cite{westerhout_plasmon_2018}. This code evaluates Eq.~\eqref{eqn-Pi_pz} for a given single-particle Hamiltonian at a given temperature $T$ and damping $\eta$. In all our calculations, the temperature was set to $k_\mathrm{B} T = 0.0256$ eV and damping was $\eta = 0.001$ eV. Compared to Ref.~\cite{westerhout_plasmon_2018} we applied two notable optimizations. First of all, we reduced the computational load by taking the sparsity of Eq.~\eqref{eqn-Eps_pz} at finite temperatures into account. This reduced the effective algorithmic complexity from $\mathcal{O}(N^4)$ to $\mathcal{O}(N^{3.13})$ where $N$ is the number of lattice sites. Furthermore, we run the computations on Graphics Processing Units (GPUs) which are much better at dense linear algebra than CPUs. All together, we could evaluate Eq.~\eqref{eqn-Pi_pz} for a given $\omega$ in less than 30 seconds on an NVIDIA V100, whereas for a comparable system size it took more than 24 hours in Ref.~\cite{westerhout_plasmon_2018}, thus achieving $\times 3000$ speedup.

    \bibliography{TBLGPlasmons.bib}

\begin{thebibliography}{65}%
\makeatletter
\providecommand \@ifxundefined [1]{%
 \@ifx{#1\undefined}
}%
\providecommand \@ifnum [1]{%
 \ifnum #1\expandafter \@firstoftwo
 \else \expandafter \@secondoftwo
 \fi
}%
\providecommand \@ifx [1]{%
 \ifx #1\expandafter \@firstoftwo
 \else \expandafter \@secondoftwo
 \fi
}%
\providecommand \natexlab [1]{#1}%
\providecommand \enquote  [1]{``#1''}%
\providecommand \bibnamefont  [1]{#1}%
\providecommand \bibfnamefont [1]{#1}%
\providecommand \citenamefont [1]{#1}%
\providecommand \href@noop [0]{\@secondoftwo}%
\providecommand \href [0]{\begingroup \@sanitize@url \@href}%
\providecommand \@href[1]{\@@startlink{#1}\@@href}%
\providecommand \@@href[1]{\endgroup#1\@@endlink}%
\providecommand \@sanitize@url [0]{\catcode `\\12\catcode `\$12\catcode
  `\&12\catcode `\#12\catcode `\^12\catcode `\_12\catcode `\%12\relax}%
\providecommand \@@startlink[1]{}%
\providecommand \@@endlink[0]{}%
\providecommand \url  [0]{\begingroup\@sanitize@url \@url }%
\providecommand \@url [1]{\endgroup\@href {#1}{\urlprefix }}%
\providecommand \urlprefix  [0]{URL }%
\providecommand \Eprint [0]{\href }%
\providecommand \doibase [0]{http://dx.doi.org/}%
\providecommand \selectlanguage [0]{\@gobble}%
\providecommand \bibinfo  [0]{\@secondoftwo}%
\providecommand \bibfield  [0]{\@secondoftwo}%
\providecommand \translation [1]{[#1]}%
\providecommand \BibitemOpen [0]{}%
\providecommand \bibitemStop [0]{}%
\providecommand \bibitemNoStop [0]{.\EOS\space}%
\providecommand \EOS [0]{\spacefactor3000\relax}%
\providecommand \BibitemShut  [1]{\csname bibitem#1\endcsname}%
\let\auto@bib@innerbib\@empty
\bibitem [{\citenamefont {Su{\'a}rez~Morell}\ \emph {et~al.}(2010)\citenamefont
  {Su{\'a}rez~Morell}, \citenamefont {Correa}, \citenamefont {Vargas},
  \citenamefont {Pacheco},\ and\ \citenamefont
  {Barticevic}}]{suarez_morell_flat_2010}%
  \BibitemOpen
  \bibfield  {author} {\bibinfo {author} {\bibfnamefont {E.}~\bibnamefont
  {Su{\'a}rez~Morell}}, \bibinfo {author} {\bibfnamefont {J.~D.}\ \bibnamefont
  {Correa}}, \bibinfo {author} {\bibfnamefont {P.}~\bibnamefont {Vargas}},
  \bibinfo {author} {\bibfnamefont {M.}~\bibnamefont {Pacheco}}, \ and\
  \bibinfo {author} {\bibfnamefont {Z.}~\bibnamefont {Barticevic}},\ }\href
  {\doibase 10.1103/PhysRevB.82.121407} {\bibfield  {journal} {\bibinfo
  {journal} {Physical Review B}\ }\textbf {\bibinfo {volume} {82}},\ \bibinfo
  {pages} {121407} (\bibinfo {year} {2010})}\BibitemShut {NoStop}%
\bibitem [{\citenamefont {Bistritzer}\ and\ \citenamefont
  {MacDonald}(2011)}]{bistritzer_moire_2011}%
  \BibitemOpen
  \bibfield  {author} {\bibinfo {author} {\bibfnamefont {R.}~\bibnamefont
  {Bistritzer}}\ and\ \bibinfo {author} {\bibfnamefont {A.~H.}\ \bibnamefont
  {MacDonald}},\ }\href {\doibase 10.1073/pnas.1108174108} {\bibfield
  {journal} {\bibinfo  {journal} {PNAS}\ }\textbf {\bibinfo {volume} {108}},\
  \bibinfo {pages} {12233} (\bibinfo {year} {2011})}\BibitemShut {NoStop}%
\bibitem [{\citenamefont {Li}\ \emph {et~al.}(2010)\citenamefont {Li},
  \citenamefont {Luican}, \citenamefont {Lopes~dos Santos}, \citenamefont
  {Castro~Neto}, \citenamefont {Reina}, \citenamefont {Kong},\ and\
  \citenamefont {Andrei}}]{li_observation_2010}%
  \BibitemOpen
  \bibfield  {author} {\bibinfo {author} {\bibfnamefont {G.}~\bibnamefont
  {Li}}, \bibinfo {author} {\bibfnamefont {A.}~\bibnamefont {Luican}}, \bibinfo
  {author} {\bibfnamefont {J.~M.~B.}\ \bibnamefont {Lopes~dos Santos}},
  \bibinfo {author} {\bibfnamefont {A.~H.}\ \bibnamefont {Castro~Neto}},
  \bibinfo {author} {\bibfnamefont {A.}~\bibnamefont {Reina}}, \bibinfo
  {author} {\bibfnamefont {J.}~\bibnamefont {Kong}}, \ and\ \bibinfo {author}
  {\bibfnamefont {E.~Y.}\ \bibnamefont {Andrei}},\ }\href {\doibase
  10.1038/nphys1463} {\bibfield  {journal} {\bibinfo  {journal} {Nature
  Physics}\ }\textbf {\bibinfo {volume} {6}},\ \bibinfo {pages} {109} (\bibinfo
  {year} {2010})}\BibitemShut {NoStop}%
\bibitem [{\citenamefont {Cao}\ \emph {et~al.}(2018{\natexlab{a}})\citenamefont
  {Cao}, \citenamefont {Fatemi}, \citenamefont {Demir}, \citenamefont {Fang},
  \citenamefont {Tomarken}, \citenamefont {Luo}, \citenamefont
  {Sanchez-Yamagishi}, \citenamefont {Watanabe}, \citenamefont {Taniguchi},
  \citenamefont {Kaxiras}, \citenamefont {Ashoori},\ and\ \citenamefont
  {Jarillo-Herrero}}]{cao_correlated_2018}%
  \BibitemOpen
  \bibfield  {author} {\bibinfo {author} {\bibfnamefont {Y.}~\bibnamefont
  {Cao}}, \bibinfo {author} {\bibfnamefont {V.}~\bibnamefont {Fatemi}},
  \bibinfo {author} {\bibfnamefont {A.}~\bibnamefont {Demir}}, \bibinfo
  {author} {\bibfnamefont {S.}~\bibnamefont {Fang}}, \bibinfo {author}
  {\bibfnamefont {S.~L.}\ \bibnamefont {Tomarken}}, \bibinfo {author}
  {\bibfnamefont {J.~Y.}\ \bibnamefont {Luo}}, \bibinfo {author} {\bibfnamefont
  {J.~D.}\ \bibnamefont {Sanchez-Yamagishi}}, \bibinfo {author} {\bibfnamefont
  {K.}~\bibnamefont {Watanabe}}, \bibinfo {author} {\bibfnamefont
  {T.}~\bibnamefont {Taniguchi}}, \bibinfo {author} {\bibfnamefont
  {E.}~\bibnamefont {Kaxiras}}, \bibinfo {author} {\bibfnamefont {R.~C.}\
  \bibnamefont {Ashoori}}, \ and\ \bibinfo {author} {\bibfnamefont
  {P.}~\bibnamefont {Jarillo-Herrero}},\ }\href {\doibase 10.1038/nature26154}
  {\bibfield  {journal} {\bibinfo  {journal} {Nature}\ }\textbf {\bibinfo
  {volume} {556}},\ \bibinfo {pages} {80} (\bibinfo {year}
  {2018}{\natexlab{a}})}\BibitemShut {NoStop}%
\bibitem [{\citenamefont {Cao}\ \emph {et~al.}(2018{\natexlab{b}})\citenamefont
  {Cao}, \citenamefont {Fatemi}, \citenamefont {Fang}, \citenamefont
  {Watanabe}, \citenamefont {Taniguchi}, \citenamefont {Kaxiras},\ and\
  \citenamefont {Jarillo-Herrero}}]{cao_unconventional_2018}%
  \BibitemOpen
  \bibfield  {author} {\bibinfo {author} {\bibfnamefont {Y.}~\bibnamefont
  {Cao}}, \bibinfo {author} {\bibfnamefont {V.}~\bibnamefont {Fatemi}},
  \bibinfo {author} {\bibfnamefont {S.}~\bibnamefont {Fang}}, \bibinfo {author}
  {\bibfnamefont {K.}~\bibnamefont {Watanabe}}, \bibinfo {author}
  {\bibfnamefont {T.}~\bibnamefont {Taniguchi}}, \bibinfo {author}
  {\bibfnamefont {E.}~\bibnamefont {Kaxiras}}, \ and\ \bibinfo {author}
  {\bibfnamefont {P.}~\bibnamefont {Jarillo-Herrero}},\ }\href {\doibase
  10.1038/nature26160} {\bibfield  {journal} {\bibinfo  {journal} {Nature}\
  }\textbf {\bibinfo {volume} {556}},\ \bibinfo {pages} {43} (\bibinfo {year}
  {2018}{\natexlab{b}})}\BibitemShut {NoStop}%
\bibitem [{\citenamefont {Yankowitz}\ \emph {et~al.}(2019)\citenamefont
  {Yankowitz}, \citenamefont {Chen}, \citenamefont {Polshyn}, \citenamefont
  {Zhang}, \citenamefont {Watanabe}, \citenamefont {Taniguchi}, \citenamefont
  {Graf}, \citenamefont {Young},\ and\ \citenamefont
  {Dean}}]{yankowitz_tuning_2019}%
  \BibitemOpen
  \bibfield  {author} {\bibinfo {author} {\bibfnamefont {M.}~\bibnamefont
  {Yankowitz}}, \bibinfo {author} {\bibfnamefont {S.}~\bibnamefont {Chen}},
  \bibinfo {author} {\bibfnamefont {H.}~\bibnamefont {Polshyn}}, \bibinfo
  {author} {\bibfnamefont {Y.}~\bibnamefont {Zhang}}, \bibinfo {author}
  {\bibfnamefont {K.}~\bibnamefont {Watanabe}}, \bibinfo {author}
  {\bibfnamefont {T.}~\bibnamefont {Taniguchi}}, \bibinfo {author}
  {\bibfnamefont {D.}~\bibnamefont {Graf}}, \bibinfo {author} {\bibfnamefont
  {A.~F.}\ \bibnamefont {Young}}, \ and\ \bibinfo {author} {\bibfnamefont
  {C.~R.}\ \bibnamefont {Dean}},\ }\href {\doibase 10.1126/science.aav1910}
  {\bibfield  {journal} {\bibinfo  {journal} {Science}\ }\textbf {\bibinfo
  {volume} {363}},\ \bibinfo {pages} {1059} (\bibinfo {year}
  {2019})}\BibitemShut {NoStop}%
\bibitem [{\citenamefont {Sharpe}\ \emph {et~al.}(2019)\citenamefont {Sharpe},
  \citenamefont {Fox}, \citenamefont {Barnard}, \citenamefont {Finney},
  \citenamefont {Watanabe}, \citenamefont {Taniguchi}, \citenamefont
  {Kastner},\ and\ \citenamefont {Goldhaber-Gordon}}]{sharpe_emergent_2019}%
  \BibitemOpen
  \bibfield  {author} {\bibinfo {author} {\bibfnamefont {A.~L.}\ \bibnamefont
  {Sharpe}}, \bibinfo {author} {\bibfnamefont {E.~J.}\ \bibnamefont {Fox}},
  \bibinfo {author} {\bibfnamefont {A.~W.}\ \bibnamefont {Barnard}}, \bibinfo
  {author} {\bibfnamefont {J.}~\bibnamefont {Finney}}, \bibinfo {author}
  {\bibfnamefont {K.}~\bibnamefont {Watanabe}}, \bibinfo {author}
  {\bibfnamefont {T.}~\bibnamefont {Taniguchi}}, \bibinfo {author}
  {\bibfnamefont {M.~A.}\ \bibnamefont {Kastner}}, \ and\ \bibinfo {author}
  {\bibfnamefont {D.}~\bibnamefont {Goldhaber-Gordon}},\ }\href {\doibase
  10.1126/science.aaw3780} {\bibfield  {journal} {\bibinfo  {journal}
  {Science}\ }\textbf {\bibinfo {volume} {365}},\ \bibinfo {pages} {605}
  (\bibinfo {year} {2019})}\BibitemShut {NoStop}%
\bibitem [{\citenamefont {Wu}\ \emph {et~al.}(2017)\citenamefont {Wu},
  \citenamefont {Lovorn},\ and\ \citenamefont
  {MacDonald}}]{wu_topological_2017}%
  \BibitemOpen
  \bibfield  {author} {\bibinfo {author} {\bibfnamefont {F.}~\bibnamefont
  {Wu}}, \bibinfo {author} {\bibfnamefont {T.}~\bibnamefont {Lovorn}}, \ and\
  \bibinfo {author} {\bibfnamefont {A.~H.}\ \bibnamefont {MacDonald}},\ }\href
  {\doibase 10.1103/PhysRevLett.118.147401} {\bibfield  {journal} {\bibinfo
  {journal} {Physical Review Letters}\ }\textbf {\bibinfo {volume} {118}},\
  \bibinfo {pages} {147401} (\bibinfo {year} {2017})}\BibitemShut {NoStop}%
\bibitem [{\citenamefont {Danovich}\ \emph {et~al.}(2018)\citenamefont
  {Danovich}, \citenamefont {Ruiz-Tijerina}, \citenamefont {Hunt},
  \citenamefont {Szyniszewski}, \citenamefont {Drummond},\ and\ \citenamefont
  {Fal'ko}}]{danovich_localized_2018}%
  \BibitemOpen
  \bibfield  {author} {\bibinfo {author} {\bibfnamefont {M.}~\bibnamefont
  {Danovich}}, \bibinfo {author} {\bibfnamefont {D.~A.}\ \bibnamefont
  {Ruiz-Tijerina}}, \bibinfo {author} {\bibfnamefont {R.~J.}\ \bibnamefont
  {Hunt}}, \bibinfo {author} {\bibfnamefont {M.}~\bibnamefont {Szyniszewski}},
  \bibinfo {author} {\bibfnamefont {N.~D.}\ \bibnamefont {Drummond}}, \ and\
  \bibinfo {author} {\bibfnamefont {V.~I.}\ \bibnamefont {Fal'ko}},\ }\href
  {\doibase 10.1103/PhysRevB.97.195452} {\bibfield  {journal} {\bibinfo
  {journal} {Physical Review B}\ }\textbf {\bibinfo {volume} {97}},\ \bibinfo
  {pages} {195452} (\bibinfo {year} {2018})}\BibitemShut {NoStop}%
\bibitem [{\citenamefont {Wu}\ \emph {et~al.}(2018)\citenamefont {Wu},
  \citenamefont {Lovorn}, \citenamefont {Tutuc},\ and\ \citenamefont
  {MacDonald}}]{wu_hubbard_2018}%
  \BibitemOpen
  \bibfield  {author} {\bibinfo {author} {\bibfnamefont {F.}~\bibnamefont
  {Wu}}, \bibinfo {author} {\bibfnamefont {T.}~\bibnamefont {Lovorn}}, \bibinfo
  {author} {\bibfnamefont {E.}~\bibnamefont {Tutuc}}, \ and\ \bibinfo {author}
  {\bibfnamefont {A.~H.}\ \bibnamefont {MacDonald}},\ }\href {\doibase
  10.1103/PhysRevLett.121.026402} {\bibfield  {journal} {\bibinfo  {journal}
  {Physical Review Letters}\ }\textbf {\bibinfo {volume} {121}},\ \bibinfo
  {pages} {026402} (\bibinfo {year} {2018})}\BibitemShut {NoStop}%
\bibitem [{\citenamefont {Brem}\ \emph {et~al.}(2020)\citenamefont {Brem},
  \citenamefont {Linder{\"a}lv}, \citenamefont {Erhart},\ and\ \citenamefont
  {Malic}}]{brem_tunable_2020}%
  \BibitemOpen
  \bibfield  {author} {\bibinfo {author} {\bibfnamefont {S.}~\bibnamefont
  {Brem}}, \bibinfo {author} {\bibfnamefont {C.}~\bibnamefont {Linder{\"a}lv}},
  \bibinfo {author} {\bibfnamefont {P.}~\bibnamefont {Erhart}}, \ and\ \bibinfo
  {author} {\bibfnamefont {E.}~\bibnamefont {Malic}},\ }\href {\doibase
  10.1021/acs.nanolett.0c03019} {\bibfield  {journal} {\bibinfo  {journal}
  {Nano Letters}\ }\textbf {\bibinfo {volume} {20}},\ \bibinfo {pages} {8534}
  (\bibinfo {year} {2020})}\BibitemShut {NoStop}%
\bibitem [{\citenamefont {Choi}\ \emph {et~al.}(2021)\citenamefont {Choi},
  \citenamefont {Florian}, \citenamefont {Steinhoff}, \citenamefont {Erben},
  \citenamefont {Tran}, \citenamefont {Kim}, \citenamefont {Sun}, \citenamefont
  {Quan}, \citenamefont {Claassen}, \citenamefont {Majumder}, \citenamefont
  {Hollingsworth}, \citenamefont {Taniguchi}, \citenamefont {Watanabe},
  \citenamefont {Ueno}, \citenamefont {Singh}, \citenamefont {Moody},
  \citenamefont {Jahnke},\ and\ \citenamefont {Li}}]{choi_twist_2021}%
  \BibitemOpen
  \bibfield  {author} {\bibinfo {author} {\bibfnamefont {J.}~\bibnamefont
  {Choi}}, \bibinfo {author} {\bibfnamefont {M.}~\bibnamefont {Florian}},
  \bibinfo {author} {\bibfnamefont {A.}~\bibnamefont {Steinhoff}}, \bibinfo
  {author} {\bibfnamefont {D.}~\bibnamefont {Erben}}, \bibinfo {author}
  {\bibfnamefont {K.}~\bibnamefont {Tran}}, \bibinfo {author} {\bibfnamefont
  {D.~S.}\ \bibnamefont {Kim}}, \bibinfo {author} {\bibfnamefont
  {L.}~\bibnamefont {Sun}}, \bibinfo {author} {\bibfnamefont {J.}~\bibnamefont
  {Quan}}, \bibinfo {author} {\bibfnamefont {R.}~\bibnamefont {Claassen}},
  \bibinfo {author} {\bibfnamefont {S.}~\bibnamefont {Majumder}}, \bibinfo
  {author} {\bibfnamefont {J.~A.}\ \bibnamefont {Hollingsworth}}, \bibinfo
  {author} {\bibfnamefont {T.}~\bibnamefont {Taniguchi}}, \bibinfo {author}
  {\bibfnamefont {K.}~\bibnamefont {Watanabe}}, \bibinfo {author}
  {\bibfnamefont {K.}~\bibnamefont {Ueno}}, \bibinfo {author} {\bibfnamefont
  {A.}~\bibnamefont {Singh}}, \bibinfo {author} {\bibfnamefont
  {G.}~\bibnamefont {Moody}}, \bibinfo {author} {\bibfnamefont
  {F.}~\bibnamefont {Jahnke}}, \ and\ \bibinfo {author} {\bibfnamefont
  {X.}~\bibnamefont {Li}},\ }\href {\doibase 10.1103/PhysRevLett.126.047401}
  {\bibfield  {journal} {\bibinfo  {journal} {Physical Review Letters}\
  }\textbf {\bibinfo {volume} {126}},\ \bibinfo {pages} {047401} (\bibinfo
  {year} {2021})}\BibitemShut {NoStop}%
\bibitem [{\citenamefont {Tran}\ \emph {et~al.}(2019)\citenamefont {Tran},
  \citenamefont {Moody}, \citenamefont {Wu}, \citenamefont {Lu}, \citenamefont
  {Choi}, \citenamefont {Kim}, \citenamefont {Rai}, \citenamefont {Sanchez},
  \citenamefont {Quan}, \citenamefont {Singh}, \citenamefont {Embley},
  \citenamefont {Zepeda}, \citenamefont {Campbell}, \citenamefont {Autry},
  \citenamefont {Taniguchi}, \citenamefont {Watanabe}, \citenamefont {Lu},
  \citenamefont {Banerjee}, \citenamefont {Silverman}, \citenamefont {Kim},
  \citenamefont {Tutuc}, \citenamefont {Yang}, \citenamefont {MacDonald},\ and\
  \citenamefont {Li}}]{tran_evidence_2019}%
  \BibitemOpen
  \bibfield  {author} {\bibinfo {author} {\bibfnamefont {K.}~\bibnamefont
  {Tran}}, \bibinfo {author} {\bibfnamefont {G.}~\bibnamefont {Moody}},
  \bibinfo {author} {\bibfnamefont {F.}~\bibnamefont {Wu}}, \bibinfo {author}
  {\bibfnamefont {X.}~\bibnamefont {Lu}}, \bibinfo {author} {\bibfnamefont
  {J.}~\bibnamefont {Choi}}, \bibinfo {author} {\bibfnamefont {K.}~\bibnamefont
  {Kim}}, \bibinfo {author} {\bibfnamefont {A.}~\bibnamefont {Rai}}, \bibinfo
  {author} {\bibfnamefont {D.~A.}\ \bibnamefont {Sanchez}}, \bibinfo {author}
  {\bibfnamefont {J.}~\bibnamefont {Quan}}, \bibinfo {author} {\bibfnamefont
  {A.}~\bibnamefont {Singh}}, \bibinfo {author} {\bibfnamefont
  {J.}~\bibnamefont {Embley}}, \bibinfo {author} {\bibfnamefont
  {A.}~\bibnamefont {Zepeda}}, \bibinfo {author} {\bibfnamefont
  {M.}~\bibnamefont {Campbell}}, \bibinfo {author} {\bibfnamefont
  {T.}~\bibnamefont {Autry}}, \bibinfo {author} {\bibfnamefont
  {T.}~\bibnamefont {Taniguchi}}, \bibinfo {author} {\bibfnamefont
  {K.}~\bibnamefont {Watanabe}}, \bibinfo {author} {\bibfnamefont
  {N.}~\bibnamefont {Lu}}, \bibinfo {author} {\bibfnamefont {S.~K.}\
  \bibnamefont {Banerjee}}, \bibinfo {author} {\bibfnamefont {K.~L.}\
  \bibnamefont {Silverman}}, \bibinfo {author} {\bibfnamefont {S.}~\bibnamefont
  {Kim}}, \bibinfo {author} {\bibfnamefont {E.}~\bibnamefont {Tutuc}}, \bibinfo
  {author} {\bibfnamefont {L.}~\bibnamefont {Yang}}, \bibinfo {author}
  {\bibfnamefont {A.~H.}\ \bibnamefont {MacDonald}}, \ and\ \bibinfo {author}
  {\bibfnamefont {X.}~\bibnamefont {Li}},\ }\href {\doibase
  10.1038/s41586-019-0975-z} {\bibfield  {journal} {\bibinfo  {journal}
  {Nature}\ }\textbf {\bibinfo {volume} {567}},\ \bibinfo {pages} {71}
  (\bibinfo {year} {2019})}\BibitemShut {NoStop}%
\bibitem [{\citenamefont {Seyler}\ \emph {et~al.}(2019)\citenamefont {Seyler},
  \citenamefont {Rivera}, \citenamefont {Yu}, \citenamefont {Wilson},
  \citenamefont {Ray}, \citenamefont {Mandrus}, \citenamefont {Yan},
  \citenamefont {Yao},\ and\ \citenamefont {Xu}}]{seyler_signatures_2019}%
  \BibitemOpen
  \bibfield  {author} {\bibinfo {author} {\bibfnamefont {K.~L.}\ \bibnamefont
  {Seyler}}, \bibinfo {author} {\bibfnamefont {P.}~\bibnamefont {Rivera}},
  \bibinfo {author} {\bibfnamefont {H.}~\bibnamefont {Yu}}, \bibinfo {author}
  {\bibfnamefont {N.~P.}\ \bibnamefont {Wilson}}, \bibinfo {author}
  {\bibfnamefont {E.~L.}\ \bibnamefont {Ray}}, \bibinfo {author} {\bibfnamefont
  {D.~G.}\ \bibnamefont {Mandrus}}, \bibinfo {author} {\bibfnamefont
  {J.}~\bibnamefont {Yan}}, \bibinfo {author} {\bibfnamefont {W.}~\bibnamefont
  {Yao}}, \ and\ \bibinfo {author} {\bibfnamefont {X.}~\bibnamefont {Xu}},\
  }\href {\doibase 10.1038/s41586-019-0957-1} {\bibfield  {journal} {\bibinfo
  {journal} {Nature}\ }\textbf {\bibinfo {volume} {567}},\ \bibinfo {pages}
  {66} (\bibinfo {year} {2019})}\BibitemShut {NoStop}%
\bibitem [{\citenamefont {Zhang}\ \emph {et~al.}(2020)\citenamefont {Zhang},
  \citenamefont {Zhang}, \citenamefont {Wu}, \citenamefont {Wang},
  \citenamefont {Gogna}, \citenamefont {Hou}, \citenamefont {Watanabe},
  \citenamefont {Taniguchi}, \citenamefont {Kulkarni}, \citenamefont {Kuo},
  \citenamefont {Forrest},\ and\ \citenamefont
  {Deng}}]{zhang_twist-angle_2020}%
  \BibitemOpen
  \bibfield  {author} {\bibinfo {author} {\bibfnamefont {L.}~\bibnamefont
  {Zhang}}, \bibinfo {author} {\bibfnamefont {Z.}~\bibnamefont {Zhang}},
  \bibinfo {author} {\bibfnamefont {F.}~\bibnamefont {Wu}}, \bibinfo {author}
  {\bibfnamefont {D.}~\bibnamefont {Wang}}, \bibinfo {author} {\bibfnamefont
  {R.}~\bibnamefont {Gogna}}, \bibinfo {author} {\bibfnamefont
  {S.}~\bibnamefont {Hou}}, \bibinfo {author} {\bibfnamefont {K.}~\bibnamefont
  {Watanabe}}, \bibinfo {author} {\bibfnamefont {T.}~\bibnamefont {Taniguchi}},
  \bibinfo {author} {\bibfnamefont {K.}~\bibnamefont {Kulkarni}}, \bibinfo
  {author} {\bibfnamefont {T.}~\bibnamefont {Kuo}}, \bibinfo {author}
  {\bibfnamefont {S.~R.}\ \bibnamefont {Forrest}}, \ and\ \bibinfo {author}
  {\bibfnamefont {H.}~\bibnamefont {Deng}},\ }\href {\doibase
  10.1038/s41467-020-19466-6} {\bibfield  {journal} {\bibinfo  {journal}
  {Nature Communications}\ }\textbf {\bibinfo {volume} {11}},\ \bibinfo {pages}
  {5888} (\bibinfo {year} {2020})}\BibitemShut {NoStop}%
\bibitem [{\citenamefont {Li}\ \emph {et~al.}(2021)\citenamefont {Li},
  \citenamefont {Li}, \citenamefont {Naik}, \citenamefont {Xie}, \citenamefont
  {Li}, \citenamefont {Wang}, \citenamefont {Regan}, \citenamefont {Wang},
  \citenamefont {Zhao}, \citenamefont {Zhao}, \citenamefont {Kahn},
  \citenamefont {Yumigeta}, \citenamefont {Blei}, \citenamefont {Taniguchi},
  \citenamefont {Watanabe}, \citenamefont {Tongay}, \citenamefont {Zettl},
  \citenamefont {Louie}, \citenamefont {Wang},\ and\ \citenamefont
  {Crommie}}]{li_imaging_2021}%
  \BibitemOpen
  \bibfield  {author} {\bibinfo {author} {\bibfnamefont {H.}~\bibnamefont
  {Li}}, \bibinfo {author} {\bibfnamefont {S.}~\bibnamefont {Li}}, \bibinfo
  {author} {\bibfnamefont {M.~H.}\ \bibnamefont {Naik}}, \bibinfo {author}
  {\bibfnamefont {J.}~\bibnamefont {Xie}}, \bibinfo {author} {\bibfnamefont
  {X.}~\bibnamefont {Li}}, \bibinfo {author} {\bibfnamefont {J.}~\bibnamefont
  {Wang}}, \bibinfo {author} {\bibfnamefont {E.}~\bibnamefont {Regan}},
  \bibinfo {author} {\bibfnamefont {D.}~\bibnamefont {Wang}}, \bibinfo {author}
  {\bibfnamefont {W.}~\bibnamefont {Zhao}}, \bibinfo {author} {\bibfnamefont
  {S.}~\bibnamefont {Zhao}}, \bibinfo {author} {\bibfnamefont {S.}~\bibnamefont
  {Kahn}}, \bibinfo {author} {\bibfnamefont {K.}~\bibnamefont {Yumigeta}},
  \bibinfo {author} {\bibfnamefont {M.}~\bibnamefont {Blei}}, \bibinfo {author}
  {\bibfnamefont {T.}~\bibnamefont {Taniguchi}}, \bibinfo {author}
  {\bibfnamefont {K.}~\bibnamefont {Watanabe}}, \bibinfo {author}
  {\bibfnamefont {S.}~\bibnamefont {Tongay}}, \bibinfo {author} {\bibfnamefont
  {A.}~\bibnamefont {Zettl}}, \bibinfo {author} {\bibfnamefont {S.~G.}\
  \bibnamefont {Louie}}, \bibinfo {author} {\bibfnamefont {F.}~\bibnamefont
  {Wang}}, \ and\ \bibinfo {author} {\bibfnamefont {M.~F.}\ \bibnamefont
  {Crommie}},\ }\href {\doibase 10.1038/s41563-021-00923-6} {\bibfield
  {journal} {\bibinfo  {journal} {Nature Materials}\ }\textbf {\bibinfo
  {volume} {20}},\ \bibinfo {pages} {945} (\bibinfo {year} {2021})}\BibitemShut
  {NoStop}%
\bibitem [{\citenamefont {Angeli}\ and\ \citenamefont
  {MacDonald}(2021)}]{angeli__2021}%
  \BibitemOpen
  \bibfield  {author} {\bibinfo {author} {\bibfnamefont {M.}~\bibnamefont
  {Angeli}}\ and\ \bibinfo {author} {\bibfnamefont {A.~H.}\ \bibnamefont
  {MacDonald}},\ }\href {\doibase 10.1073/pnas.2021826118} {\bibfield
  {journal} {\bibinfo  {journal} {Proceedings of the National Academy of
  Sciences}\ }\textbf {\bibinfo {volume} {118}} (\bibinfo {year} {2021}),\
  10.1073/pnas.2021826118}\BibitemShut {NoStop}%
\bibitem [{\citenamefont {Goodwin}\ \emph
  {et~al.}(2019{\natexlab{a}})\citenamefont {Goodwin}, \citenamefont
  {Corsetti}, \citenamefont {Mostofi},\ and\ \citenamefont
  {Lischner}}]{goodwin_twist-angle_2019}%
  \BibitemOpen
  \bibfield  {author} {\bibinfo {author} {\bibfnamefont {Z.~A.~H.}\
  \bibnamefont {Goodwin}}, \bibinfo {author} {\bibfnamefont {F.}~\bibnamefont
  {Corsetti}}, \bibinfo {author} {\bibfnamefont {A.~A.}\ \bibnamefont
  {Mostofi}}, \ and\ \bibinfo {author} {\bibfnamefont {J.}~\bibnamefont
  {Lischner}},\ }\href {\doibase 10.1103/PhysRevB.100.121106} {\bibfield
  {journal} {\bibinfo  {journal} {Physical Review B}\ }\textbf {\bibinfo
  {volume} {100}},\ \bibinfo {pages} {121106} (\bibinfo {year}
  {2019}{\natexlab{a}})}\BibitemShut {NoStop}%
\bibitem [{\citenamefont {Cea}\ and\ \citenamefont
  {Guinea}(2020)}]{cea_band_2020}%
  \BibitemOpen
  \bibfield  {author} {\bibinfo {author} {\bibfnamefont {T.}~\bibnamefont
  {Cea}}\ and\ \bibinfo {author} {\bibfnamefont {F.}~\bibnamefont {Guinea}},\
  }\href {\doibase 10.1103/PhysRevB.102.045107} {\bibfield  {journal} {\bibinfo
   {journal} {Physical Review B}\ }\textbf {\bibinfo {volume} {102}},\ \bibinfo
  {pages} {045107} (\bibinfo {year} {2020})}\BibitemShut {NoStop}%
\bibitem [{\citenamefont {Bernevig}\ \emph {et~al.}(2021)\citenamefont
  {Bernevig}, \citenamefont {Song}, \citenamefont {Regnault},\ and\
  \citenamefont {Lian}}]{bernevig_twisted_2021}%
  \BibitemOpen
  \bibfield  {author} {\bibinfo {author} {\bibfnamefont {B.~A.}\ \bibnamefont
  {Bernevig}}, \bibinfo {author} {\bibfnamefont {Z.-D.}\ \bibnamefont {Song}},
  \bibinfo {author} {\bibfnamefont {N.}~\bibnamefont {Regnault}}, \ and\
  \bibinfo {author} {\bibfnamefont {B.}~\bibnamefont {Lian}},\ }\href {\doibase
  10.1103/PhysRevB.103.205413} {\bibfield  {journal} {\bibinfo  {journal}
  {Physical Review B}\ }\textbf {\bibinfo {volume} {103}},\ \bibinfo {pages}
  {205413} (\bibinfo {year} {2021})}\BibitemShut {NoStop}%
\bibitem [{\citenamefont {Liu}\ \emph {et~al.}(2021)\citenamefont {Liu},
  \citenamefont {Wang}, \citenamefont {Watanabe}, \citenamefont {Taniguchi},
  \citenamefont {Vafek},\ and\ \citenamefont {Li}}]{liu_tuning_2021}%
  \BibitemOpen
  \bibfield  {author} {\bibinfo {author} {\bibfnamefont {X.}~\bibnamefont
  {Liu}}, \bibinfo {author} {\bibfnamefont {Z.}~\bibnamefont {Wang}}, \bibinfo
  {author} {\bibfnamefont {K.}~\bibnamefont {Watanabe}}, \bibinfo {author}
  {\bibfnamefont {T.}~\bibnamefont {Taniguchi}}, \bibinfo {author}
  {\bibfnamefont {O.}~\bibnamefont {Vafek}}, \ and\ \bibinfo {author}
  {\bibfnamefont {J.~I.~A.}\ \bibnamefont {Li}},\ }\href {\doibase
  10.1126/science.abb8754} {\bibfield  {journal} {\bibinfo  {journal}
  {Science}\ }\textbf {\bibinfo {volume} {371}},\ \bibinfo {pages} {1261}
  (\bibinfo {year} {2021})}\BibitemShut {NoStop}%
\bibitem [{\citenamefont {Pizarro}\ \emph {et~al.}(2019)\citenamefont
  {Pizarro}, \citenamefont {R{\"o}sner}, \citenamefont {Thomale}, \citenamefont
  {Valent{\'i}},\ and\ \citenamefont {Wehling}}]{pizarro_internal_2019}%
  \BibitemOpen
  \bibfield  {author} {\bibinfo {author} {\bibfnamefont {J.~M.}\ \bibnamefont
  {Pizarro}}, \bibinfo {author} {\bibfnamefont {M.}~\bibnamefont {R{\"o}sner}},
  \bibinfo {author} {\bibfnamefont {R.}~\bibnamefont {Thomale}}, \bibinfo
  {author} {\bibfnamefont {R.}~\bibnamefont {Valent{\'i}}}, \ and\ \bibinfo
  {author} {\bibfnamefont {T.~O.}\ \bibnamefont {Wehling}},\ }\href {\doibase
  10.1103/PhysRevB.100.161102} {\bibfield  {journal} {\bibinfo  {journal}
  {Physical Review B}\ }\textbf {\bibinfo {volume} {100}},\ \bibinfo {pages}
  {161102} (\bibinfo {year} {2019})}\BibitemShut {NoStop}%
\bibitem [{\citenamefont {Goodwin}\ \emph
  {et~al.}(2019{\natexlab{b}})\citenamefont {Goodwin}, \citenamefont
  {Corsetti}, \citenamefont {Mostofi},\ and\ \citenamefont
  {Lischner}}]{goodwin_attractive_2019}%
  \BibitemOpen
  \bibfield  {author} {\bibinfo {author} {\bibfnamefont {Z.~A.~H.}\
  \bibnamefont {Goodwin}}, \bibinfo {author} {\bibfnamefont {F.}~\bibnamefont
  {Corsetti}}, \bibinfo {author} {\bibfnamefont {A.~A.}\ \bibnamefont
  {Mostofi}}, \ and\ \bibinfo {author} {\bibfnamefont {J.}~\bibnamefont
  {Lischner}},\ }\href {\doibase 10.1103/PhysRevB.100.235424} {\bibfield
  {journal} {\bibinfo  {journal} {Physical Review B}\ }\textbf {\bibinfo
  {volume} {100}},\ \bibinfo {pages} {235424} (\bibinfo {year}
  {2019}{\natexlab{b}})}\BibitemShut {NoStop}%
\bibitem [{\citenamefont {Sunku}\ \emph {et~al.}(2018)\citenamefont {Sunku},
  \citenamefont {Ni}, \citenamefont {Jiang}, \citenamefont {Yoo}, \citenamefont
  {Sternbach}, \citenamefont {McLeod}, \citenamefont {Stauber}, \citenamefont
  {Xiong}, \citenamefont {Taniguchi}, \citenamefont {Watanabe}, \citenamefont
  {Kim}, \citenamefont {Fogler},\ and\ \citenamefont
  {Basov}}]{sunku_photonic_2018}%
  \BibitemOpen
  \bibfield  {author} {\bibinfo {author} {\bibfnamefont {S.~S.}\ \bibnamefont
  {Sunku}}, \bibinfo {author} {\bibfnamefont {G.~X.}\ \bibnamefont {Ni}},
  \bibinfo {author} {\bibfnamefont {B.~Y.}\ \bibnamefont {Jiang}}, \bibinfo
  {author} {\bibfnamefont {H.}~\bibnamefont {Yoo}}, \bibinfo {author}
  {\bibfnamefont {A.}~\bibnamefont {Sternbach}}, \bibinfo {author}
  {\bibfnamefont {A.~S.}\ \bibnamefont {McLeod}}, \bibinfo {author}
  {\bibfnamefont {T.}~\bibnamefont {Stauber}}, \bibinfo {author} {\bibfnamefont
  {L.}~\bibnamefont {Xiong}}, \bibinfo {author} {\bibfnamefont
  {T.}~\bibnamefont {Taniguchi}}, \bibinfo {author} {\bibfnamefont
  {K.}~\bibnamefont {Watanabe}}, \bibinfo {author} {\bibfnamefont
  {P.}~\bibnamefont {Kim}}, \bibinfo {author} {\bibfnamefont {M.~M.}\
  \bibnamefont {Fogler}}, \ and\ \bibinfo {author} {\bibfnamefont {D.~N.}\
  \bibnamefont {Basov}},\ }\href {\doibase 10.1126/science.aau5144} {\bibfield
  {journal} {\bibinfo  {journal} {Science}\ }\textbf {\bibinfo {volume}
  {362}},\ \bibinfo {pages} {1153} (\bibinfo {year} {2018})}\BibitemShut
  {NoStop}%
\bibitem [{\citenamefont {Lewandowski}\ and\ \citenamefont
  {Levitov}(2019)}]{lewandowski_intrinsically_2019}%
  \BibitemOpen
  \bibfield  {author} {\bibinfo {author} {\bibfnamefont {C.}~\bibnamefont
  {Lewandowski}}\ and\ \bibinfo {author} {\bibfnamefont {L.}~\bibnamefont
  {Levitov}},\ }\href {\doibase 10.1073/pnas.1909069116} {\bibfield  {journal}
  {\bibinfo  {journal} {Proceedings of the National Academy of Sciences}\
  }\textbf {\bibinfo {volume} {116}},\ \bibinfo {pages} {20869} (\bibinfo
  {year} {2019})}\BibitemShut {NoStop}%
\bibitem [{\citenamefont {Hesp}\ \emph {et~al.}(2019)\citenamefont {Hesp},
  \citenamefont {Torre}, \citenamefont {Rodan-Legrain}, \citenamefont
  {Novelli}, \citenamefont {Cao}, \citenamefont {Carr}, \citenamefont {Fang},
  \citenamefont {Stepanov}, \citenamefont {Barcons-Ruiz}, \citenamefont
  {Herzig-Sheinfux}, \citenamefont {Watanabe}, \citenamefont {Taniguchi},
  \citenamefont {Efetov}, \citenamefont {Kaxiras}, \citenamefont
  {Jarillo-Herrero}, \citenamefont {Polini},\ and\ \citenamefont
  {Koppens}}]{hesp_collective_2019}%
  \BibitemOpen
  \bibfield  {author} {\bibinfo {author} {\bibfnamefont {N.~C.~H.}\
  \bibnamefont {Hesp}}, \bibinfo {author} {\bibfnamefont {I.}~\bibnamefont
  {Torre}}, \bibinfo {author} {\bibfnamefont {D.}~\bibnamefont
  {Rodan-Legrain}}, \bibinfo {author} {\bibfnamefont {P.}~\bibnamefont
  {Novelli}}, \bibinfo {author} {\bibfnamefont {Y.}~\bibnamefont {Cao}},
  \bibinfo {author} {\bibfnamefont {S.}~\bibnamefont {Carr}}, \bibinfo {author}
  {\bibfnamefont {S.}~\bibnamefont {Fang}}, \bibinfo {author} {\bibfnamefont
  {P.}~\bibnamefont {Stepanov}}, \bibinfo {author} {\bibfnamefont
  {D.}~\bibnamefont {Barcons-Ruiz}}, \bibinfo {author} {\bibfnamefont
  {H.}~\bibnamefont {Herzig-Sheinfux}}, \bibinfo {author} {\bibfnamefont
  {K.}~\bibnamefont {Watanabe}}, \bibinfo {author} {\bibfnamefont
  {T.}~\bibnamefont {Taniguchi}}, \bibinfo {author} {\bibfnamefont {D.~K.}\
  \bibnamefont {Efetov}}, \bibinfo {author} {\bibfnamefont {E.}~\bibnamefont
  {Kaxiras}}, \bibinfo {author} {\bibfnamefont {P.}~\bibnamefont
  {Jarillo-Herrero}}, \bibinfo {author} {\bibfnamefont {M.}~\bibnamefont
  {Polini}}, \ and\ \bibinfo {author} {\bibfnamefont {F.~H.~L.}\ \bibnamefont
  {Koppens}},\ }\href {http://arxiv.org/abs/1910.07893} {\bibfield  {journal}
  {\bibinfo  {journal} {arXiv:1910.07893 [cond-mat]}\ } (\bibinfo {year}
  {2019})}\BibitemShut {NoStop}%
\bibitem [{\citenamefont {Novelli}\ \emph {et~al.}(2020)\citenamefont
  {Novelli}, \citenamefont {Torre}, \citenamefont {Koppens}, \citenamefont
  {Taddei},\ and\ \citenamefont {Polini}}]{novelli_optical_2020}%
  \BibitemOpen
  \bibfield  {author} {\bibinfo {author} {\bibfnamefont {P.}~\bibnamefont
  {Novelli}}, \bibinfo {author} {\bibfnamefont {I.}~\bibnamefont {Torre}},
  \bibinfo {author} {\bibfnamefont {F.~H.~L.}\ \bibnamefont {Koppens}},
  \bibinfo {author} {\bibfnamefont {F.}~\bibnamefont {Taddei}}, \ and\ \bibinfo
  {author} {\bibfnamefont {M.}~\bibnamefont {Polini}},\ }\href {\doibase
  10.1103/PhysRevB.102.125403} {\bibfield  {journal} {\bibinfo  {journal}
  {Physical Review B}\ }\textbf {\bibinfo {volume} {102}},\ \bibinfo {pages}
  {125403} (\bibinfo {year} {2020})}\BibitemShut {NoStop}%
\bibitem [{\citenamefont {Hu}\ \emph {et~al.}(2017)\citenamefont {Hu},
  \citenamefont {Das}, \citenamefont {Luan}, \citenamefont {Chung},
  \citenamefont {Chen},\ and\ \citenamefont {Fei}}]{hu_real-space_2017}%
  \BibitemOpen
  \bibfield  {author} {\bibinfo {author} {\bibfnamefont {F.}~\bibnamefont
  {Hu}}, \bibinfo {author} {\bibfnamefont {S.~R.}\ \bibnamefont {Das}},
  \bibinfo {author} {\bibfnamefont {Y.}~\bibnamefont {Luan}}, \bibinfo {author}
  {\bibfnamefont {T.-F.}\ \bibnamefont {Chung}}, \bibinfo {author}
  {\bibfnamefont {Y.~P.}\ \bibnamefont {Chen}}, \ and\ \bibinfo {author}
  {\bibfnamefont {Z.}~\bibnamefont {Fei}},\ }\href {\doibase
  10.1103/PhysRevLett.119.247402} {\bibfield  {journal} {\bibinfo  {journal}
  {Physical Review Letters}\ }\textbf {\bibinfo {volume} {119}},\ \bibinfo
  {pages} {247402} (\bibinfo {year} {2017})}\BibitemShut {NoStop}%
\bibitem [{\citenamefont {Katsnelson}\ and\ \citenamefont
  {Trefilov}(1985)}]{katsnelson_anomalies_1985}%
  \BibitemOpen
  \bibfield  {author} {\bibinfo {author} {\bibfnamefont {M.~I.}\ \bibnamefont
  {Katsnelson}}\ and\ \bibinfo {author} {\bibfnamefont {A.~V.}\ \bibnamefont
  {Trefilov}},\ }\href@noop {} {\bibfield  {journal} {\bibinfo  {journal} {JETP
  Letters}\ }\textbf {\bibinfo {volume} {42}},\ \bibinfo {pages} {485}
  (\bibinfo {year} {1985})}\BibitemShut {NoStop}%
\bibitem [{\citenamefont {Irkhin}\ \emph {et~al.}(2002)\citenamefont {Irkhin},
  \citenamefont {Katanin},\ and\ \citenamefont
  {Katsnelson}}]{irkhin_robustness_2002}%
  \BibitemOpen
  \bibfield  {author} {\bibinfo {author} {\bibfnamefont {V.~Y.}\ \bibnamefont
  {Irkhin}}, \bibinfo {author} {\bibfnamefont {A.~A.}\ \bibnamefont {Katanin}},
  \ and\ \bibinfo {author} {\bibfnamefont {M.~I.}\ \bibnamefont {Katsnelson}},\
  }\href {\doibase 10.1103/PhysRevLett.89.076401} {\bibfield  {journal}
  {\bibinfo  {journal} {Physical Review Letters}\ }\textbf {\bibinfo {volume}
  {89}},\ \bibinfo {pages} {076401} (\bibinfo {year} {2002})}\BibitemShut
  {NoStop}%
\bibitem [{\citenamefont {Stepanov}\ \emph {et~al.}(2021)\citenamefont
  {Stepanov}, \citenamefont {Harkov}, \citenamefont {R{\"o}sner}, \citenamefont
  {Lichtenstein}, \citenamefont {Katsnelson},\ and\ \citenamefont
  {Rudenko}}]{stepanov_coexisting_2021}%
  \BibitemOpen
  \bibfield  {author} {\bibinfo {author} {\bibfnamefont {E.~A.}\ \bibnamefont
  {Stepanov}}, \bibinfo {author} {\bibfnamefont {V.}~\bibnamefont {Harkov}},
  \bibinfo {author} {\bibfnamefont {M.}~\bibnamefont {R{\"o}sner}}, \bibinfo
  {author} {\bibfnamefont {A.~I.}\ \bibnamefont {Lichtenstein}}, \bibinfo
  {author} {\bibfnamefont {M.~I.}\ \bibnamefont {Katsnelson}}, \ and\ \bibinfo
  {author} {\bibfnamefont {A.~N.}\ \bibnamefont {Rudenko}},\ }\href
  {http://arxiv.org/abs/2107.01132} {\bibfield  {journal} {\bibinfo  {journal}
  {arXiv:2107.01132 [cond-mat]}\ } (\bibinfo {year} {2021})}\BibitemShut
  {NoStop}%
\bibitem [{\citenamefont {Shi}\ \emph {et~al.}(2020)\citenamefont {Shi},
  \citenamefont {Zhan}, \citenamefont {Qi}, \citenamefont {Huang},
  \citenamefont {Veen}, \citenamefont {Silva-Guill{\'e}n}, \citenamefont
  {Zhang}, \citenamefont {Li}, \citenamefont {Xie}, \citenamefont {Ji},
  \citenamefont {Katsnelson}, \citenamefont {Yuan}, \citenamefont {Qin},\ and\
  \citenamefont {Zhang}}]{shi_large-area_2020}%
  \BibitemOpen
  \bibfield  {author} {\bibinfo {author} {\bibfnamefont {H.}~\bibnamefont
  {Shi}}, \bibinfo {author} {\bibfnamefont {Z.}~\bibnamefont {Zhan}}, \bibinfo
  {author} {\bibfnamefont {Z.}~\bibnamefont {Qi}}, \bibinfo {author}
  {\bibfnamefont {K.}~\bibnamefont {Huang}}, \bibinfo {author} {\bibfnamefont
  {E.~v.}\ \bibnamefont {Veen}}, \bibinfo {author} {\bibfnamefont {J.~{\'A}.}\
  \bibnamefont {Silva-Guill{\'e}n}}, \bibinfo {author} {\bibfnamefont
  {R.}~\bibnamefont {Zhang}}, \bibinfo {author} {\bibfnamefont
  {P.}~\bibnamefont {Li}}, \bibinfo {author} {\bibfnamefont {K.}~\bibnamefont
  {Xie}}, \bibinfo {author} {\bibfnamefont {H.}~\bibnamefont {Ji}}, \bibinfo
  {author} {\bibfnamefont {M.~I.}\ \bibnamefont {Katsnelson}}, \bibinfo
  {author} {\bibfnamefont {S.}~\bibnamefont {Yuan}}, \bibinfo {author}
  {\bibfnamefont {S.}~\bibnamefont {Qin}}, \ and\ \bibinfo {author}
  {\bibfnamefont {Z.}~\bibnamefont {Zhang}},\ }\href {\doibase
  10.1038/s41467-019-14207-w} {\bibfield  {journal} {\bibinfo  {journal}
  {Nature Communications}\ }\textbf {\bibinfo {volume} {11}},\ \bibinfo {pages}
  {371} (\bibinfo {year} {2020})}\BibitemShut {NoStop}%
\bibitem [{\citenamefont {Gornostyrev}\ and\ \citenamefont
  {Katsnelson}(2020)}]{gornostyrev_origin_2020}%
  \BibitemOpen
  \bibfield  {author} {\bibinfo {author} {\bibfnamefont {Y.~N.}\ \bibnamefont
  {Gornostyrev}}\ and\ \bibinfo {author} {\bibfnamefont {M.~I.}\ \bibnamefont
  {Katsnelson}},\ }\href {\doibase 10.1103/PhysRevB.102.085428} {\bibfield
  {journal} {\bibinfo  {journal} {Physical Review B}\ }\textbf {\bibinfo
  {volume} {102}},\ \bibinfo {pages} {085428} (\bibinfo {year}
  {2020})}\BibitemShut {NoStop}%
\bibitem [{\citenamefont {van Wijk}\ \emph {et~al.}(2015)\citenamefont {van
  Wijk}, \citenamefont {Schuring}, \citenamefont {Katsnelson},\ and\
  \citenamefont {Fasolino}}]{wijk_relaxation_2015}%
  \BibitemOpen
  \bibfield  {author} {\bibinfo {author} {\bibfnamefont {M.~M.}\ \bibnamefont
  {van Wijk}}, \bibinfo {author} {\bibfnamefont {A.}~\bibnamefont {Schuring}},
  \bibinfo {author} {\bibfnamefont {M.~I.}\ \bibnamefont {Katsnelson}}, \ and\
  \bibinfo {author} {\bibfnamefont {A.}~\bibnamefont {Fasolino}},\ }\href
  {\doibase 10.1088/2053-1583/2/3/034010} {\bibfield  {journal} {\bibinfo
  {journal} {2D Materials}\ }\textbf {\bibinfo {volume} {2}},\ \bibinfo {pages}
  {034010} (\bibinfo {year} {2015})}\BibitemShut {NoStop}%
\bibitem [{\citenamefont {Vonsovsky}\ and\ \citenamefont
  {Katsnelson}(1989)}]{vonsovsky_quantum_1989}%
  \BibitemOpen
  \bibfield  {author} {\bibinfo {author} {\bibfnamefont {S.~V.}\ \bibnamefont
  {Vonsovsky}}\ and\ \bibinfo {author} {\bibfnamefont {M.~I.}\ \bibnamefont
  {Katsnelson}},\ }\href {https://www.springer.com/gp/book/9783642501661}
  {\emph {\bibinfo {title} {Quantum {Solid}-{State} {Physics}}}},\ Springer
  {Series} in {Solid}-{State} {Sciences}\ (\bibinfo  {publisher}
  {Springer-Verlag},\ \bibinfo {address} {Berlin Heidelberg},\ \bibinfo {year}
  {1989})\BibitemShut {NoStop}%
\bibitem [{\citenamefont {Giuliani}\ and\ \citenamefont
  {Vignale}(2005)}]{giuliani_quantum_2005}%
  \BibitemOpen
  \bibfield  {author} {\bibinfo {author} {\bibfnamefont {G.}~\bibnamefont
  {Giuliani}}\ and\ \bibinfo {author} {\bibfnamefont {G.}~\bibnamefont
  {Vignale}},\ }\href {\doibase 10.1017/CBO9780511619915} {\emph {\bibinfo
  {title} {Quantum {Theory} of the {Electron} {Liquid}}}}\ (\bibinfo
  {publisher} {Cambridge University Press},\ \bibinfo {address} {Cambridge},\
  \bibinfo {year} {2005})\BibitemShut {NoStop}%
\bibitem [{\citenamefont {Wang}\ \emph {et~al.}(2015)\citenamefont {Wang},
  \citenamefont {Christensen}, \citenamefont {Jauho}, \citenamefont {Thygesen},
  \citenamefont {Wubs},\ and\ \citenamefont {Mortensen}}]{wang_plasmonic_2015}%
  \BibitemOpen
  \bibfield  {author} {\bibinfo {author} {\bibfnamefont {W.}~\bibnamefont
  {Wang}}, \bibinfo {author} {\bibfnamefont {T.}~\bibnamefont {Christensen}},
  \bibinfo {author} {\bibfnamefont {A.-P.}\ \bibnamefont {Jauho}}, \bibinfo
  {author} {\bibfnamefont {K.~S.}\ \bibnamefont {Thygesen}}, \bibinfo {author}
  {\bibfnamefont {M.}~\bibnamefont {Wubs}}, \ and\ \bibinfo {author}
  {\bibfnamefont {N.~A.}\ \bibnamefont {Mortensen}},\ }\href {\doibase
  10.1038/srep09535} {\bibfield  {journal} {\bibinfo  {journal} {Scientific
  Reports}\ }\textbf {\bibinfo {volume} {5}},\ \bibinfo {pages} {9535}
  (\bibinfo {year} {2015})}\BibitemShut {NoStop}%
\bibitem [{\citenamefont {Westerhout}\ \emph {et~al.}(2018)\citenamefont
  {Westerhout}, \citenamefont {van Veen}, \citenamefont {Katsnelson},\ and\
  \citenamefont {Yuan}}]{westerhout_plasmon_2018}%
  \BibitemOpen
  \bibfield  {author} {\bibinfo {author} {\bibfnamefont {T.}~\bibnamefont
  {Westerhout}}, \bibinfo {author} {\bibfnamefont {E.}~\bibnamefont {van
  Veen}}, \bibinfo {author} {\bibfnamefont {M.~I.}\ \bibnamefont {Katsnelson}},
  \ and\ \bibinfo {author} {\bibfnamefont {S.}~\bibnamefont {Yuan}},\ }\href
  {\doibase 10.1103/PhysRevB.97.205434} {\bibfield  {journal} {\bibinfo
  {journal} {Physical Review B}\ }\textbf {\bibinfo {volume} {97}},\ \bibinfo
  {pages} {205434} (\bibinfo {year} {2018})}\BibitemShut {NoStop}%
\bibitem [{\citenamefont {Jiang}\ \emph {et~al.}(2021)\citenamefont {Jiang},
  \citenamefont {Haas},\ and\ \citenamefont
  {R{\"o}sner}}]{jiang_plasmonic_2021}%
  \BibitemOpen
  \bibfield  {author} {\bibinfo {author} {\bibfnamefont {Z.}~\bibnamefont
  {Jiang}}, \bibinfo {author} {\bibfnamefont {S.}~\bibnamefont {Haas}}, \ and\
  \bibinfo {author} {\bibfnamefont {M.}~\bibnamefont {R{\"o}sner}},\ }\href
  {\doibase 10.1088/2053-1583/abfedd} {\bibfield  {journal} {\bibinfo
  {journal} {2D Materials}\ }\textbf {\bibinfo {volume} {8}},\ \bibinfo {pages}
  {035037} (\bibinfo {year} {2021})}\BibitemShut {NoStop}%
\bibitem [{\citenamefont {Guinea}\ and\ \citenamefont
  {Walet}(2019)}]{guinea_continuum_2019}%
  \BibitemOpen
  \bibfield  {author} {\bibinfo {author} {\bibfnamefont {F.}~\bibnamefont
  {Guinea}}\ and\ \bibinfo {author} {\bibfnamefont {N.~R.}\ \bibnamefont
  {Walet}},\ }\href {\doibase 10.1103/PhysRevB.99.205134} {\bibfield  {journal}
  {\bibinfo  {journal} {Physical Review B}\ }\textbf {\bibinfo {volume} {99}},\
  \bibinfo {pages} {205134} (\bibinfo {year} {2019})}\BibitemShut {NoStop}%
\bibitem [{\citenamefont {van Loon}\ \emph {et~al.}(2021)\citenamefont {van
  Loon}, \citenamefont {R{\"o}sner}, \citenamefont {Katsnelson},\ and\
  \citenamefont {Wehling}}]{van_loon_random_2021}%
  \BibitemOpen
  \bibfield  {author} {\bibinfo {author} {\bibfnamefont {E.~G. C.~P.}\
  \bibnamefont {van Loon}}, \bibinfo {author} {\bibfnamefont {M.}~\bibnamefont
  {R{\"o}sner}}, \bibinfo {author} {\bibfnamefont {M.~I.}\ \bibnamefont
  {Katsnelson}}, \ and\ \bibinfo {author} {\bibfnamefont {T.~O.}\ \bibnamefont
  {Wehling}},\ }\href {http://arxiv.org/abs/2103.04419} {\bibfield  {journal}
  {\bibinfo  {journal} {arXiv:2103.04419 [cond-mat]}\ } (\bibinfo {year}
  {2021})}\BibitemShut {NoStop}%
\bibitem [{\citenamefont {Aryasetiawan}\ \emph {et~al.}(2004)\citenamefont
  {Aryasetiawan}, \citenamefont {Imada}, \citenamefont {Georges}, \citenamefont
  {Kotliar}, \citenamefont {Biermann},\ and\ \citenamefont
  {Lichtenstein}}]{cRPA}%
  \BibitemOpen
  \bibfield  {author} {\bibinfo {author} {\bibfnamefont {F.}~\bibnamefont
  {Aryasetiawan}}, \bibinfo {author} {\bibfnamefont {M.}~\bibnamefont {Imada}},
  \bibinfo {author} {\bibfnamefont {A.}~\bibnamefont {Georges}}, \bibinfo
  {author} {\bibfnamefont {G.}~\bibnamefont {Kotliar}}, \bibinfo {author}
  {\bibfnamefont {S.}~\bibnamefont {Biermann}}, \ and\ \bibinfo {author}
  {\bibfnamefont {A.~I.}\ \bibnamefont {Lichtenstein}},\ }\href {\doibase
  10.1103/PhysRevB.70.195104} {\bibfield  {journal} {\bibinfo  {journal} {Phys.
  Rev. B}\ }\textbf {\bibinfo {volume} {70}},\ \bibinfo {pages} {195104}
  (\bibinfo {year} {2004})}\BibitemShut {NoStop}%
\bibitem [{\citenamefont {Keldysh}(1979)}]{keldysh_coulomb_1979}%
  \BibitemOpen
  \bibfield  {author} {\bibinfo {author} {\bibfnamefont {L.~V.}\ \bibnamefont
  {Keldysh}},\ }\href {https://ui.adsabs.harvard.edu/abs/1979JETPL..29..658K}
  {\bibfield  {journal} {\bibinfo  {journal} {Soviet Journal of Experimental
  and Theoretical Physics Letters}\ }\textbf {\bibinfo {volume} {29}},\
  \bibinfo {pages} {658} (\bibinfo {year} {1979})}\BibitemShut {NoStop}%
\bibitem [{\citenamefont {Jena}\ and\ \citenamefont
  {Konar}(2007)}]{jena_enhancement_2007}%
  \BibitemOpen
  \bibfield  {author} {\bibinfo {author} {\bibfnamefont {D.}~\bibnamefont
  {Jena}}\ and\ \bibinfo {author} {\bibfnamefont {A.}~\bibnamefont {Konar}},\
  }\href {\doibase 10.1103/PhysRevLett.98.136805} {\bibfield  {journal}
  {\bibinfo  {journal} {Physical Review Letters}\ }\textbf {\bibinfo {volume}
  {98}},\ \bibinfo {pages} {136805} (\bibinfo {year} {2007})}\BibitemShut
  {NoStop}%
\bibitem [{\citenamefont {Emelyanenko}\ and\ \citenamefont
  {Boinovich}(2008)}]{emelyanenko_effect_2008}%
  \BibitemOpen
  \bibfield  {author} {\bibinfo {author} {\bibfnamefont {A.}~\bibnamefont
  {Emelyanenko}}\ and\ \bibinfo {author} {\bibfnamefont {L.}~\bibnamefont
  {Boinovich}},\ }\href {\doibase 10.1088/0953-8984/20/49/494227} {\bibfield
  {journal} {\bibinfo  {journal} {Journal of Physics: Condensed Matter}\
  }\textbf {\bibinfo {volume} {20}},\ \bibinfo {pages} {494227} (\bibinfo
  {year} {2008})}\BibitemShut {NoStop}%
\bibitem [{\citenamefont {R{\"o}sner}\ \emph {et~al.}(2015)\citenamefont
  {R{\"o}sner}, \citenamefont {{\c S}a{\c s}{\i}o{\u g}lu}, \citenamefont
  {Friedrich}, \citenamefont {Bl{\"u}gel},\ and\ \citenamefont
  {Wehling}}]{rosner_wannier_2015}%
  \BibitemOpen
  \bibfield  {author} {\bibinfo {author} {\bibfnamefont {M.}~\bibnamefont
  {R{\"o}sner}}, \bibinfo {author} {\bibfnamefont {E.}~\bibnamefont {{\c S}a{\c
  s}{\i}o{\u g}lu}}, \bibinfo {author} {\bibfnamefont {C.}~\bibnamefont
  {Friedrich}}, \bibinfo {author} {\bibfnamefont {S.}~\bibnamefont
  {Bl{\"u}gel}}, \ and\ \bibinfo {author} {\bibfnamefont {T.~O.}\ \bibnamefont
  {Wehling}},\ }\href {\doibase 10.1103/PhysRevB.92.085102} {\bibfield
  {journal} {\bibinfo  {journal} {Physical Review B}\ }\textbf {\bibinfo
  {volume} {92}},\ \bibinfo {pages} {085102} (\bibinfo {year}
  {2015})}\BibitemShut {NoStop}%
\bibitem [{\citenamefont {Katsnelson}(2006)}]{katsnelson_nonlinear_2006}%
  \BibitemOpen
  \bibfield  {author} {\bibinfo {author} {\bibfnamefont {M.~I.}\ \bibnamefont
  {Katsnelson}},\ }\href {\doibase 10.1103/PhysRevB.74.201401} {\bibfield
  {journal} {\bibinfo  {journal} {Physical Review B}\ }\textbf {\bibinfo
  {volume} {74}},\ \bibinfo {pages} {201401} (\bibinfo {year}
  {2006})}\BibitemShut {NoStop}%
\bibitem [{\citenamefont {Zheliuk}\ \emph {et~al.}(2019)\citenamefont
  {Zheliuk}, \citenamefont {Lu}, \citenamefont {Chen}, \citenamefont {Yumin},
  \citenamefont {Golightly},\ and\ \citenamefont
  {Ye}}]{zheliuk_josephson_2019}%
  \BibitemOpen
  \bibfield  {author} {\bibinfo {author} {\bibfnamefont {O.}~\bibnamefont
  {Zheliuk}}, \bibinfo {author} {\bibfnamefont {J.~M.}\ \bibnamefont {Lu}},
  \bibinfo {author} {\bibfnamefont {Q.~H.}\ \bibnamefont {Chen}}, \bibinfo
  {author} {\bibfnamefont {A.~A.~E.}\ \bibnamefont {Yumin}}, \bibinfo {author}
  {\bibfnamefont {S.}~\bibnamefont {Golightly}}, \ and\ \bibinfo {author}
  {\bibfnamefont {J.~T.}\ \bibnamefont {Ye}},\ }\href {\doibase
  10.1038/s41565-019-0564-1} {\bibfield  {journal} {\bibinfo  {journal} {Nature
  Nanotechnology}\ }\textbf {\bibinfo {volume} {14}},\ \bibinfo {pages} {1123}
  (\bibinfo {year} {2019})}\BibitemShut {NoStop}%
\bibitem [{\citenamefont {da~Jornada}\ \emph {et~al.}(2020)\citenamefont
  {da~Jornada}, \citenamefont {Xian}, \citenamefont {Rubio},\ and\
  \citenamefont {Louie}}]{da_jornada_universal_2020}%
  \BibitemOpen
  \bibfield  {author} {\bibinfo {author} {\bibfnamefont {F.~H.}\ \bibnamefont
  {da~Jornada}}, \bibinfo {author} {\bibfnamefont {L.}~\bibnamefont {Xian}},
  \bibinfo {author} {\bibfnamefont {A.}~\bibnamefont {Rubio}}, \ and\ \bibinfo
  {author} {\bibfnamefont {S.~G.}\ \bibnamefont {Louie}},\ }\href {\doibase
  10.1038/s41467-020-14826-8} {\bibfield  {journal} {\bibinfo  {journal}
  {Nature Communications}\ }\textbf {\bibinfo {volume} {11}},\ \bibinfo {pages}
  {1013} (\bibinfo {year} {2020})}\BibitemShut {NoStop}%
\bibitem [{\citenamefont {Das~Sarma}\ and\ \citenamefont
  {Hwang}(1998)}]{das_sarma_plasmons_1998}%
  \BibitemOpen
  \bibfield  {author} {\bibinfo {author} {\bibfnamefont {S.}~\bibnamefont
  {Das~Sarma}}\ and\ \bibinfo {author} {\bibfnamefont {E.~H.}\ \bibnamefont
  {Hwang}},\ }\href {\doibase 10.1103/PhysRevLett.81.4216} {\bibfield
  {journal} {\bibinfo  {journal} {Physical Review Letters}\ }\textbf {\bibinfo
  {volume} {81}},\ \bibinfo {pages} {4216} (\bibinfo {year}
  {1998})}\BibitemShut {NoStop}%
\bibitem [{\citenamefont {Hwang}\ and\ \citenamefont
  {Das~Sarma}(2009)}]{hwang_plasmon_2009}%
  \BibitemOpen
  \bibfield  {author} {\bibinfo {author} {\bibfnamefont {E.~H.}\ \bibnamefont
  {Hwang}}\ and\ \bibinfo {author} {\bibfnamefont {S.}~\bibnamefont
  {Das~Sarma}},\ }\href {\doibase 10.1103/PhysRevB.80.205405} {\bibfield
  {journal} {\bibinfo  {journal} {Physical Review B}\ }\textbf {\bibinfo
  {volume} {80}},\ \bibinfo {pages} {205405} (\bibinfo {year}
  {2009})}\BibitemShut {NoStop}%
\bibitem [{\citenamefont {Rold{\'a}n}\ and\ \citenamefont
  {Brey}(2013)}]{roldan_dielectric_2013}%
  \BibitemOpen
  \bibfield  {author} {\bibinfo {author} {\bibfnamefont {R.}~\bibnamefont
  {Rold{\'a}n}}\ and\ \bibinfo {author} {\bibfnamefont {L.}~\bibnamefont
  {Brey}},\ }\href {\doibase 10.1103/PhysRevB.88.115420} {\bibfield  {journal}
  {\bibinfo  {journal} {Physical Review B}\ }\textbf {\bibinfo {volume} {88}},\
  \bibinfo {pages} {115420} (\bibinfo {year} {2013})}\BibitemShut {NoStop}%
\bibitem [{\citenamefont {Jin}\ \emph {et~al.}(2015)\citenamefont {Jin},
  \citenamefont {Rold{\'a}n}, \citenamefont {Katsnelson},\ and\ \citenamefont
  {Yuan}}]{jin_screening_2015}%
  \BibitemOpen
  \bibfield  {author} {\bibinfo {author} {\bibfnamefont {F.}~\bibnamefont
  {Jin}}, \bibinfo {author} {\bibfnamefont {R.}~\bibnamefont {Rold{\'a}n}},
  \bibinfo {author} {\bibfnamefont {M.~I.}\ \bibnamefont {Katsnelson}}, \ and\
  \bibinfo {author} {\bibfnamefont {S.}~\bibnamefont {Yuan}},\ }\href {\doibase
  10.1103/PhysRevB.92.115440} {\bibfield  {journal} {\bibinfo  {journal}
  {Physical Review B}\ }\textbf {\bibinfo {volume} {92}},\ \bibinfo {pages}
  {115440} (\bibinfo {year} {2015})}\BibitemShut {NoStop}%
\bibitem [{\citenamefont {Fei}\ \emph {et~al.}(2012)\citenamefont {Fei},
  \citenamefont {Rodin}, \citenamefont {Andreev}, \citenamefont {Bao},
  \citenamefont {McLeod}, \citenamefont {Wagner}, \citenamefont {Zhang},
  \citenamefont {Zhao}, \citenamefont {Thiemens}, \citenamefont {Dominguez},
  \citenamefont {Fogler}, \citenamefont {Neto}, \citenamefont {Lau},
  \citenamefont {Keilmann},\ and\ \citenamefont
  {Basov}}]{fei_gate-tuning_2012}%
  \BibitemOpen
  \bibfield  {author} {\bibinfo {author} {\bibfnamefont {Z.}~\bibnamefont
  {Fei}}, \bibinfo {author} {\bibfnamefont {A.~S.}\ \bibnamefont {Rodin}},
  \bibinfo {author} {\bibfnamefont {G.~O.}\ \bibnamefont {Andreev}}, \bibinfo
  {author} {\bibfnamefont {W.}~\bibnamefont {Bao}}, \bibinfo {author}
  {\bibfnamefont {A.~S.}\ \bibnamefont {McLeod}}, \bibinfo {author}
  {\bibfnamefont {M.}~\bibnamefont {Wagner}}, \bibinfo {author} {\bibfnamefont
  {L.~M.}\ \bibnamefont {Zhang}}, \bibinfo {author} {\bibfnamefont
  {Z.}~\bibnamefont {Zhao}}, \bibinfo {author} {\bibfnamefont {M.}~\bibnamefont
  {Thiemens}}, \bibinfo {author} {\bibfnamefont {G.}~\bibnamefont {Dominguez}},
  \bibinfo {author} {\bibfnamefont {M.~M.}\ \bibnamefont {Fogler}}, \bibinfo
  {author} {\bibfnamefont {A.~H.~C.}\ \bibnamefont {Neto}}, \bibinfo {author}
  {\bibfnamefont {C.~N.}\ \bibnamefont {Lau}}, \bibinfo {author} {\bibfnamefont
  {F.}~\bibnamefont {Keilmann}}, \ and\ \bibinfo {author} {\bibfnamefont
  {D.~N.}\ \bibnamefont {Basov}},\ }\href {\doibase 10.1038/nature11253}
  {\bibfield  {journal} {\bibinfo  {journal} {Nature}\ }\textbf {\bibinfo
  {volume} {487}},\ \bibinfo {pages} {82} (\bibinfo {year} {2012})}\BibitemShut
  {NoStop}%
\bibitem [{\citenamefont {Chen}\ \emph {et~al.}(2012)\citenamefont {Chen},
  \citenamefont {Badioli}, \citenamefont {Alonso-Gonz{\'a}lez}, \citenamefont
  {Thongrattanasiri}, \citenamefont {Huth}, \citenamefont {Osmond},
  \citenamefont {Spasenovi{\'c}}, \citenamefont {Centeno}, \citenamefont
  {Pesquera}, \citenamefont {Godignon}, \citenamefont {Zurutuza~Elorza},
  \citenamefont {Camara}, \citenamefont {de~Abajo}, \citenamefont
  {Hillenbrand},\ and\ \citenamefont {Koppens}}]{chen_optical_2012}%
  \BibitemOpen
  \bibfield  {author} {\bibinfo {author} {\bibfnamefont {J.}~\bibnamefont
  {Chen}}, \bibinfo {author} {\bibfnamefont {M.}~\bibnamefont {Badioli}},
  \bibinfo {author} {\bibfnamefont {P.}~\bibnamefont {Alonso-Gonz{\'a}lez}},
  \bibinfo {author} {\bibfnamefont {S.}~\bibnamefont {Thongrattanasiri}},
  \bibinfo {author} {\bibfnamefont {F.}~\bibnamefont {Huth}}, \bibinfo {author}
  {\bibfnamefont {J.}~\bibnamefont {Osmond}}, \bibinfo {author} {\bibfnamefont
  {M.}~\bibnamefont {Spasenovi{\'c}}}, \bibinfo {author} {\bibfnamefont
  {A.}~\bibnamefont {Centeno}}, \bibinfo {author} {\bibfnamefont
  {A.}~\bibnamefont {Pesquera}}, \bibinfo {author} {\bibfnamefont
  {P.}~\bibnamefont {Godignon}}, \bibinfo {author} {\bibfnamefont
  {A.}~\bibnamefont {Zurutuza~Elorza}}, \bibinfo {author} {\bibfnamefont
  {N.}~\bibnamefont {Camara}}, \bibinfo {author} {\bibfnamefont {F.~J.~G.}\
  \bibnamefont {de~Abajo}}, \bibinfo {author} {\bibfnamefont {R.}~\bibnamefont
  {Hillenbrand}}, \ and\ \bibinfo {author} {\bibfnamefont {F.~H.~L.}\
  \bibnamefont {Koppens}},\ }\href {\doibase 10.1038/nature11254} {\bibfield
  {journal} {\bibinfo  {journal} {Nature}\ }\textbf {\bibinfo {volume} {487}},\
  \bibinfo {pages} {77} (\bibinfo {year} {2012})}\BibitemShut {NoStop}%
\bibitem [{\citenamefont {Bl\"ochl}(1994)}]{paw1}%
  \BibitemOpen
  \bibfield  {author} {\bibinfo {author} {\bibfnamefont {P.~E.}\ \bibnamefont
  {Bl\"ochl}},\ }\href {\doibase 10.1103/PhysRevB.50.17953} {\bibfield
  {journal} {\bibinfo  {journal} {Phys. Rev. B}\ }\textbf {\bibinfo {volume}
  {50}},\ \bibinfo {pages} {17953} (\bibinfo {year} {1994})}\BibitemShut
  {NoStop}%
\bibitem [{\citenamefont {Kresse}\ and\ \citenamefont {Joubert}(1999)}]{paw2}%
  \BibitemOpen
  \bibfield  {author} {\bibinfo {author} {\bibfnamefont {G.}~\bibnamefont
  {Kresse}}\ and\ \bibinfo {author} {\bibfnamefont {D.}~\bibnamefont
  {Joubert}},\ }\href {https://link.aps.org/doi/10.1103/PhysRevB.59.1758}
  {\bibfield  {journal} {\bibinfo  {journal} {Phys. Rev. B}\ }\textbf {\bibinfo
  {volume} {59}},\ \bibinfo {pages} {1758} (\bibinfo {year}
  {1999})}\BibitemShut {NoStop}%
\bibitem [{\citenamefont {Kresse}\ and\ \citenamefont
  {Furthmüller}(1996)}]{Kresse1}%
  \BibitemOpen
  \bibfield  {author} {\bibinfo {author} {\bibfnamefont {G.}~\bibnamefont
  {Kresse}}\ and\ \bibinfo {author} {\bibfnamefont {J.}~\bibnamefont
  {Furthmüller}},\ }\href {\doibase 10.1016/0927-0256(96)00008-0} {\bibfield
  {journal} {\bibinfo  {journal} {Comput. Mater. Sci.}\ }\textbf {\bibinfo
  {volume} {6}},\ \bibinfo {pages} {15–50} (\bibinfo {year}
  {1996})}\BibitemShut {NoStop}%
\bibitem [{\citenamefont {Kresse}\ and\ \citenamefont
  {Furthm\"uller}(1996)}]{Kresse2}%
  \BibitemOpen
  \bibfield  {author} {\bibinfo {author} {\bibfnamefont {G.}~\bibnamefont
  {Kresse}}\ and\ \bibinfo {author} {\bibfnamefont {J.}~\bibnamefont
  {Furthm\"uller}},\ }\href {\doibase doi.org/10.1103/PhysRevB.54.11169}
  {\bibfield  {journal} {\bibinfo  {journal} {Phys. Rev. B}\ }\textbf {\bibinfo
  {volume} {54}},\ \bibinfo {pages} {11169} (\bibinfo {year}
  {1996})}\BibitemShut {NoStop}%
\bibitem [{\citenamefont {Perdew}\ \emph {et~al.}(1996)\citenamefont {Perdew},
  \citenamefont {Burke},\ and\ \citenamefont {Ernzerhof}}]{gga}%
  \BibitemOpen
  \bibfield  {author} {\bibinfo {author} {\bibfnamefont {J.~P.}\ \bibnamefont
  {Perdew}}, \bibinfo {author} {\bibfnamefont {K.}~\bibnamefont {Burke}}, \
  and\ \bibinfo {author} {\bibfnamefont {M.}~\bibnamefont {Ernzerhof}},\ }\href
  {\doibase doi.org/10.1103/PhysRevLett.77.3865} {\bibfield  {journal}
  {\bibinfo  {journal} {Phys. Rev. Lett.}\ }\textbf {\bibinfo {volume} {77}},\
  \bibinfo {pages} {3865} (\bibinfo {year} {1996})}\BibitemShut {NoStop}%
\bibitem [{\citenamefont {Marzari}\ and\ \citenamefont
  {Vanderbilt}(1997)}]{mlwf1}%
  \BibitemOpen
  \bibfield  {author} {\bibinfo {author} {\bibfnamefont {N.}~\bibnamefont
  {Marzari}}\ and\ \bibinfo {author} {\bibfnamefont {D.}~\bibnamefont
  {Vanderbilt}},\ }\href {\doibase doi.org/10.1103/PhysRevB.56.12847}
  {\bibfield  {journal} {\bibinfo  {journal} {Phys. Rev. B}\ }\textbf {\bibinfo
  {volume} {56}},\ \bibinfo {pages} {12847} (\bibinfo {year}
  {1997})}\BibitemShut {NoStop}%
\bibitem [{\citenamefont {Marzari}\ \emph {et~al.}(2012)\citenamefont
  {Marzari}, \citenamefont {Mostofi}, \citenamefont {Yates}, \citenamefont
  {Souza},\ and\ \citenamefont {Vanderbilt}}]{mlwf2}%
  \BibitemOpen
  \bibfield  {author} {\bibinfo {author} {\bibfnamefont {N.}~\bibnamefont
  {Marzari}}, \bibinfo {author} {\bibfnamefont {A.~A.}\ \bibnamefont
  {Mostofi}}, \bibinfo {author} {\bibfnamefont {J.~R.}\ \bibnamefont {Yates}},
  \bibinfo {author} {\bibfnamefont {I.}~\bibnamefont {Souza}}, \ and\ \bibinfo
  {author} {\bibfnamefont {D.}~\bibnamefont {Vanderbilt}},\ }\href {\doibase
  10.1103/RevModPhys.84.1419} {\bibfield  {journal} {\bibinfo  {journal} {Rev.
  Mod. Phys.}\ }\textbf {\bibinfo {volume} {84}},\ \bibinfo {pages} {1419}
  (\bibinfo {year} {2012})}\BibitemShut {NoStop}%
\bibitem [{\citenamefont {Mostofi}\ \emph {et~al.}(2008)\citenamefont
  {Mostofi}, \citenamefont {Yates}, \citenamefont {Lee}, \citenamefont {Souza},
  \citenamefont {Vanderbilt},\ and\ \citenamefont {Marzari}}]{wannier90}%
  \BibitemOpen
  \bibfield  {author} {\bibinfo {author} {\bibfnamefont {A.~A.}\ \bibnamefont
  {Mostofi}}, \bibinfo {author} {\bibfnamefont {J.~R.}\ \bibnamefont {Yates}},
  \bibinfo {author} {\bibfnamefont {Y.-S.}\ \bibnamefont {Lee}}, \bibinfo
  {author} {\bibfnamefont {I.}~\bibnamefont {Souza}}, \bibinfo {author}
  {\bibfnamefont {D.}~\bibnamefont {Vanderbilt}}, \ and\ \bibinfo {author}
  {\bibfnamefont {N.}~\bibnamefont {Marzari}},\ }\href {\doibase
  doi.org/10.1016/j.cpc.2007.11.016} {\bibfield  {journal} {\bibinfo  {journal}
  {Comput. Phys. Commun.}\ }\textbf {\bibinfo {volume} {178}},\ \bibinfo
  {pages} {685 } (\bibinfo {year} {2008})}\BibitemShut {NoStop}%
\bibitem [{\citenamefont {Miyake}\ and\ \citenamefont
  {Aryasetiawan}(2008)}]{CoulombU}%
  \BibitemOpen
  \bibfield  {author} {\bibinfo {author} {\bibfnamefont {T.}~\bibnamefont
  {Miyake}}\ and\ \bibinfo {author} {\bibfnamefont {F.}~\bibnamefont
  {Aryasetiawan}},\ }\href {\doibase 10.1103/PhysRevB.77.085122} {\bibfield
  {journal} {\bibinfo  {journal} {Phys. Rev. B}\ }\textbf {\bibinfo {volume}
  {77}},\ \bibinfo {pages} {085122} (\bibinfo {year} {2008})}\BibitemShut
  {NoStop}%
\bibitem [{\citenamefont {Kaltak}()}]{KaltakcRPA}%
  \BibitemOpen
  \bibfield  {author} {\bibinfo {author} {\bibfnamefont {M.}~\bibnamefont
  {Kaltak}},\ }\href {http://othes.univie.ac.at/38099/} {\enquote {\bibinfo
  {title} {Merging {GW} with {DMFT}},}\ }\bibinfo {note} {{PhD Thesis,
  University of Vienna, 2015, 231 pp.}}\BibitemShut {Stop}%
\end{thebibliography}%

\end{document}